\documentclass[%
preprint, 
superscriptaddress,
amsmath,amssymb,
aps, physrev,
]{revtex4-2}
\usepackage{xcolor}%
\usepackage{booktabs}%
\usepackage{mathtools,tikz,caption}
\DeclareRobustCommand\sampleline[1]{%
	\tikz\draw[#1] (0,0) (0,\the\dimexpr\fontdimen22\textfont2\relax)
	-- (2em,\the\dimexpr\fontdimen22\textfont2\relax);%
}
\usepackage{graphicx}% Include figure files
\usepackage{dcolumn}% Align table columns on decimal point
\usepackage{bm}% bold math
\newcommand\RF[1]{\textcolor{black}{#1}}

\begin{document}

%\preprint{APS/123-QED}

\title{Experimental study of turbulent mixing in a T-shaped mixer}% Force line breaks with \\

\author{Huixin Li} 
\affiliation{%
	Center of Applied Space Technology and Microgravity (ZARM), University of Bremen, Am Falltum 2, 28359, Bremen, Germany
}%
\affiliation{The State Key Laboratory of Nonlinear Mechanics, Institute of Mechanics, Chinese Academy of Sciences, No. 15 Beisihuanxi Road, 100190, Beijing, China}
\affiliation{School of Engineering Science, University of Chinese Academy of Sciences, No. 1 Yanqihu East Road, 101408, Beijing, China}%Lines break automatically or can be forced with \\
%\altaffiliation[Also at ]{The State Key Laboratory of Nonlinear Mechanics, Institute of Mechanics, Chinese Academy of Sciences, No. 15 Beisihuanxi Road, 100190, Beijing, China\\ School of Engineering Science, University of Chinese Academy of Sciences, No. 1 Yanqihu East Road, 100049, Beijing, China}%Lines break automatically or can be forced with \\

\author{Mohammad Mehdi Zamani Asl}%
\affiliation{%
	Center of Applied Space Technology and Microgravity (ZARM), University of Bremen, Am Falltum 2, 28359, Bremen, Germany
}%

\author{Bastian Bäuerlein}%
\affiliation{%
	Institute of Physics, University of Oldenburg, Ammerländer Heerstrasse. 114-118, 26129, Oldenburg, Germany
}%
\affiliation{%
	{ForWind-Center for Wind Energy Research, Küpkersweg 70, 26129 Oldenburg, Germany} 
}%

\author{Kerstin Avila}%
\affiliation{%
	Institute of Physics, University of Oldenburg, Ammerländer Heerstrasse. 114-118, 26129, Oldenburg, Germany
}%
\affiliation{%
	{ForWind-Center for Wind Energy Research, Küpkersweg 70, 26129 Oldenburg, Germany} 
}%

\author{Duo Xu}%
\email{duo.xu@imech.ac.cn}
\affiliation{%
	Center of Applied Space Technology and Microgravity (ZARM), University of Bremen, Am Falltum 2, 28359, Bremen, Germany
}%
\affiliation{The State Key Laboratory of Nonlinear Mechanics, Institute of Mechanics, Chinese Academy of Sciences, No. 15 Beisihuanxi Road, 100190, Beijing, China}
\affiliation{School of Engineering Science, University of Chinese Academy of Sciences, No. 1 Yanqihu East Road, 101408, Beijing, China}%Lines break automatically or can be forced with \\

\author{Marc Avila}%
\affiliation{%
	Center of Applied Space Technology and Microgravity (ZARM), University of Bremen, Am Falltum 2, 28359, Bremen, Germany
}%
%\affiliation{
%\affiliation{
%\affiliation{%


\begin{abstract}
One of the most widespread canonical devices for fluid mixing is the T-shaped mixer, in which two opposing miscible liquid streams meet at a junction and then mix along a main channel. Laminar steady and time-periodic flows in T-shaped mixers have been thoroughly studied, but turbulent flows have received much less scrutiny despite their prevalence in applications. We here introduce a novel experimental setup with a hydraulic diameter of four centimetres that enables the optical study of turbulent mixing at small scales. Using this setup, we perform two-dimensional particle image velocimetry and planar laser-induced fluorescence measurements. First, we successfully replicate characteristic flow regimes observed in micro-scale T-shaped mixers at low Reynolds numbers. We then focus on the turbulent regime and characterize the turbulent kinetic energy and dissipation along the mixing channel. Further, we measure the scalar concentration variance and its corresponding probability density function and spectra. The latter exhibits an incipient Batchelor scaling. We estimate the mechanical-to-scalar timescale ratio and examine the link between the turbulent velocity and scalar fields. The measurement data are compared with model predictions and correlations used in engineering practice, and with our own direct numerical simulations performed with a spectral-element code. 
%\begin{description}
%\end{description}
\end{abstract}

\maketitle

%\tableofcontents

\section{Introduction}
\label{intro}

Fluid mixing is an important multi-scale process in nature and in engineering applications. The T-shaped mixer, hereafter referred as T-mixer, is a canonical system widely used for fundamental research in fluid mechanics and as a testbed in chemical engineering. In its simplest and most studied form the inlets have square cross-section, and the main (outlet) channel is of rectangular shape. The macroscopic mixing properties in the laminar and transitional regimes ({Reynolds number} $Re=HU_0/\nu\lesssim500$, where $U_0$ is the mean flow speed in the inlet channels, $H$ their hydraulic diameter and $\nu$ the fluid's kinematic viscosity) are reasonably well understood. Specifically, the mixing quality has been unambiguously linked to the changes in the large-scale structures emerging in a sequence of flow instabilities, as explained in what follows (see e.g., the review \cite{camarri2020overview}). At low Reynolds numbers, $Re\lesssim120$, the two fluid streams are laminar and flow parallel to each other along the main channel. Here the mixing is purely diffusive, because the vortices in the flow are reflection symmetric and thus do not contribute to mixing the two streams \cite{hoffmann2006experimental,Bothe2006}. When the Reynolds number is increased, Kelvin--Helmholtz-like vortices emerge at the T-junction, and form a steady engulfment regime, {where the vortices in the flow are asymmetric} \cite{hoffmann2006,Bothe2006,Fani2013,mariotti2018steady}. In this regime, the contact surface of the two streams becomes strongly convoluted and leads to enhanced mixing, although the velocity and concentration fields are time-independent. {When the Reynolds number is larger than about $220$}, the engulfment regime becomes time-periodic \cite{fani2014,mariotti2018steady} and chaotic \cite{thomas2010experimental}. As the Reynolds number further increases, the engulfment regime is superseded by a symmetric regime, where the flow is time-periodic and regains the left-right reflection symmetry. Consequently, the mixing efficiency drops after this transition \cite{thomas2010mixing,schikarski2017direct}. Subsequently the flow becomes chaotic and transitions to turbulence. For $Re\gtrsim 650$, the flow in the outlet channel is already turbulent \citep{schikarski2019inflow}. We note that only a few numerical studies have explored the transition to turbulence and the turbulent regime in the outlet channel \citep{minakov2013investigation,schikarski2019inflow}. A few experiments and simulations focusing on particle precipitation in T-mixers have covered wide Reynolds number ranges, $Re\in[250,15000]$ and have demonstrated the key influence of flow regimes and mixing on the particle size distributions \cite{Gradl2006CEP, Schwarzer2006CES, Schikarski2019CET,schikarski2022quantitative}. 

In addition to the Reynolds number, the mixing dynamics is also governed by the Schmidt number $Sc=\nu/D$, where $D$ is the molecular diffusivity of the scalar. In liquid mixing, $Sc=\mathcal{O}(10^3)$, which is technically challenging in both experiments and numerical simulations due to the very small scale dynamics associated with large $Sc$. Specifically, when the flow is turbulent, the scalar mixing occurs down to the Batchelor length scale~\citep{batchelor1959small}, $\eta_b=(\nu D^2/\epsilon)^{1/4}$, where $\epsilon$ is the viscous dissipation of turbulent kinetic energy. The Batchelor scale is a factor $\sqrt{Sc}$ smaller than the smallest length scale in the velocity field,  the Kolmogorov scale $\eta_k=(\nu^3/\epsilon)^{1/4}$\citep{Pope_2000}.  Thus, in direct numerical simulations of liquid mixing, resolving the Batchelor scale demands enormous computing power \cite{gotoh_yeung_2012,buaria2021turbulence}. Specifically, for $Sc=1000$, about  a factor of $Sc^{3/2}\approx31,623$ more grid points are necessary to resolve the Batchelor scale in the scalar equation than for the Kolmogorv scale in the Navier--Stokes equations. In the experiments, measuring the fine fluctuations of the scalar field is also a technical challenge due to the limited measurement dynamic range in scale and limited signal-to-noise ratio for the weak scalar signal at small scales. 

For large $Re$ and isotropic conditions, the spectrum of the velocity field features the classical Kolmogorov $\kappa^{-5/3}$ in the inertial-convection range, followed by a sharper viscous-convection range down to the wave number ($\kappa$) corresponding to the Kolmogorov scale. In the spectrum of passive scalar fluctuations, \citet{batchelor1959small} and \citet{Kraichnan68} theoretically predicted that there exists a $\kappa^{-1}$ scaling for the length scales between the Kolmogorov and the Batchelor scales. Past evidence for this scaling law for the power spectrum is scarce and ambiguous. \citet{gibson1963universal} measured the passive scalar in grid turbulence at a point, used  Taylor's hypothesis to infer the spatial dynamics and fitted a $\kappa^{-1}$ to the resulting spectrum, which however did not have a clear scaling range. \citet{grant1968spectrum} in an ocean and tidal channel reported that the temperature fluctuations are consistent with a $\kappa^{-1}$ scaling. In a turbulent jet flow, \citet{talbot2009time} used hot-wire anemometry and Rayleigh light scattering to simultaneously measure the velocity and concentration field, in which the scalar spectra exhibits $\kappa^{-1}$ scaling law (again leveraging Taylor's frozen hypothesis). Similarly, \citet{iwano2021power} used an optical fiber laser-induced fluorescence probe with a 2.8 $\mu$m spatial resolution to investigate the power spectrum in a high Schmidt number turbulent jet. Their measured spectrum approximately follows the $\kappa^{-1}$ scaling but has a small bump. Similar results were also reported in other measurements with turbulent jets \citep[]{miller1991stochastic, miller1991reynolds, miller1996measurements}. Recently, \citet{clay2017strained} provided the evidence that the $\kappa^{-1}$ scaling region does exist and becomes better defined with increasing $Sc$ through direct numerical simulation. Despite this great progress, it appears necessary to revisit this scaling through experiments and without Taylor's frozen hypothesis. Measuring the mixing processes down to length scales as close to the Batchelor scale as possible, would help to understand the fundamental mixing physics, and subsequently to develop appropriate models for direct numerical simulations resolving the velocity field only, and for large eddy simulation (LES) and Reynolds-averaged Navier--Stokes simulation (RANS). In fact, \citet{schikarski2022quantitative} have recently shown that the uncertainty in the mixing time is the main factor hindering accurate, parameter-free predictions of reactions in mixers.

In engineering practice, RANS are commonly used for the reactor design and to predict fluid mixing, given their modest computing cost and acceptable accuracy in the flow patterns and turbulent dissipation. They are combined with  semi-empirical microscale mixing formulas to predict the mixing performance \citep{Baldyga99,Fox_2003, Liu2006AIChE}. The performance of the mixing models needs to be evaluated by experimentally measuring the bulk quantities, such as mixing efficiency, or/and the detailed dynamic process, which is particularly useful to guide reactor design. The evaluation of the model performance is essentially important for high Reynolds number flows, as evidenced in impinging jet mixers, where the experiments and the RANS simulations show noticeable discrepancies in mean velocity and turbulent kinetic energy \cite{Liu2009LabOnChip, Gavi2010CERD}. Similar model evaluations and detailed comparisons between simulation and experiment are missing for turbulent flows in T-mixers.

We here measure the scalar concentration in a T-mixer using planar laser-induced fluorescence (PLIF) and the velocity field with particle image velocimetry (PIV). Our T-mixer has large hydraulic diameter, $H=40$~{mm}, and enables optical measurements of small-scale mixing that are infeasible in the submillimeter T-mixers typically used in the literature. Our experiments cover from the laminar to the fully turbulent regime, $Re\in[100,5000]$, and provide detailed quantitative information of the mixing processes along the mixing channel. Long inlets allow the development of the flow before entering the junction and enable well-defined boundary conditions for the comparison with direct numerical simulations, which we also perform at $Re=1500$ for validation processes. The rest of the paper is structured as follows. In \S\ref{sec:method} we present our experimental setup and specify the measurement and simulation methods. PIV measurements of the velocity field are presented in \S\ref{sec:vel} and cover the inlet flow conditions, a validation of the laminar and transitional flows patterns previously reported for flows in small experimental setups and in simulations, and detailed analysis of the turbulent velocity field. The scalar mixing along the main channel is analyzed in  \S\ref{sec:sca} .  Models used in engineering practice are also assessed with our data. Our main findings are briefly summarized in \S\ref{sec:conclusion}.

\section{Methods}\label{sec:method}
\subsection{Experimental setup}
\begin{figure*}
	\centering
	\includegraphics[width=1\textwidth]{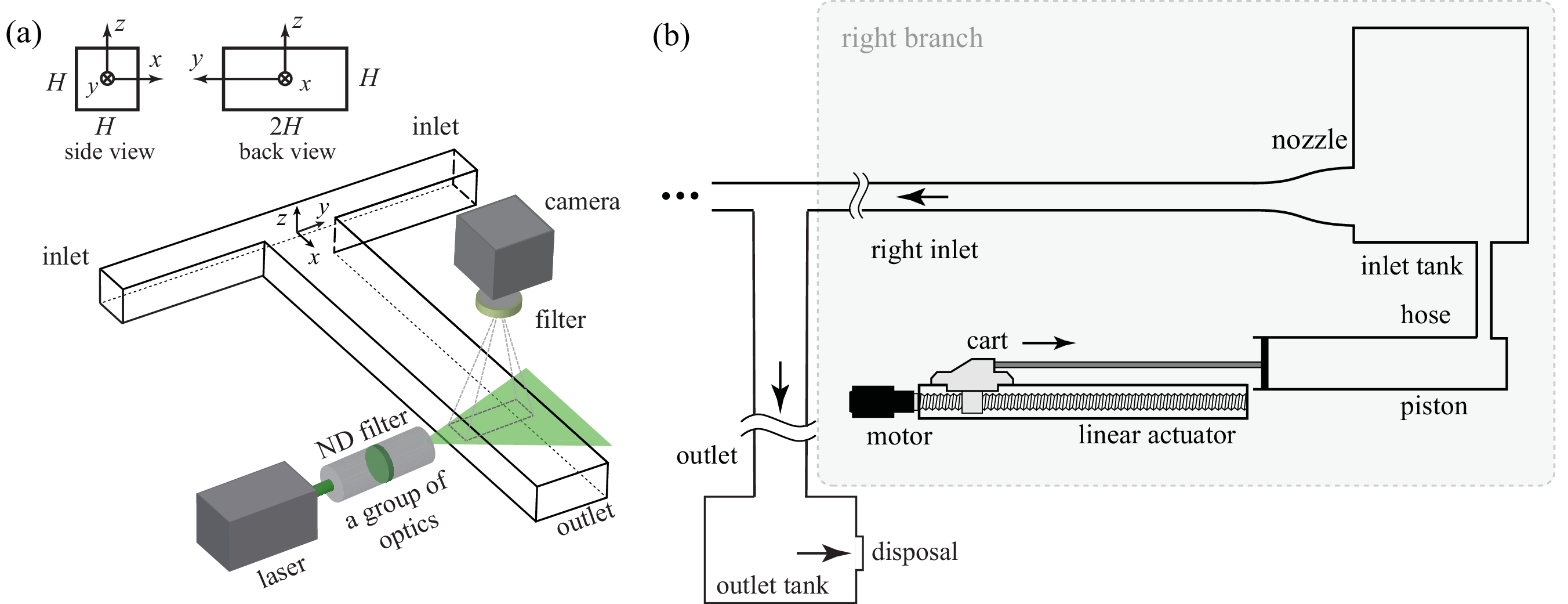}
	\caption{(Color online) (a) Three-dimensional sketch of the T-mixer setup, {and measurement hardware arrangement (i.e. PIV and PLIF). The T-mixer consists of two inlets meeting at the T-junction followed by the mixing (outlet) channel. The main measurement hardware includes a laser, a group of optics and a CMOS camera (with a camera lens and a filter). Here the green cylindrical shape means the neutral density (ND) filter and the light green cylindrical shape is the camera lens filter of PIV or PLIF.} The origin of the coordinates is set at the center of the junction. (b) Schematic illustration of the right branch of the T-mixer.}\label{fig:sketch}
\end{figure*}
The T-mixer is schematically illustrated in figure~\ref{fig:sketch}(a). {The Reynolds number is defined as $Re=HU_0/\nu$, {where $U_0$ is the mean flow speed in the inlet channels}, $H$ is the height of the channels and $\nu$ is the kinematic viscosity of water.} The T-mixer consists of two identical inlet channels, one T-shaped junction and one outlet channel. {A segment of channel is made from acrylic glass. The square cross-section for both inlets has $H = 40\pm0.25$~{mm}. Inner corners at the junction have small radii of curvature ($< 1$ mm), with negligible impact on flow behavior \cite{thomas2010mixing}. Each channel segment is 500 mm long. The channel segments are connected with custom-made adapters, sealed with O-rings, polytetrafluoroethylene tape, and silicone. The whole setup is supported by Bosch profiles and was carefully leveled.} Both inlets of the setup have a length of 3~m ($75H$), which, as shown later, allows fully developed laminar flow at the far downstream of the inlet channel ($75H$) up to $Re\lesssim1500$. The outlet channel of the T-mixer has a rectangular cross-section in dimension of $2H \times H$ (width$\times$height), and the channel is $25H$ long.  For the analysis, we use a Cartesian coordinate system with the origin positioned at the center of the T-junction, see the sketch in figure~\ref{fig:sketch}(a). The $y$ axis is along the inlet, $x$ points in the downstream direction along the mixing (outlet) channel and $z$ is perpendicular to $x-y$ plane.

For each of the inlet channels, the fluid is driven by a system composed of a motor, a linear actuator and a piston. The servo motor ({MS2N04, 570W, Bosch-Rexroth$^\circledR$}) is precisely controlled via a National Instrument card and the software {IndraWorks}. The motor is connected to the threaded shaft of the linear actuator (see figure~\ref{fig:sketch}b), the cart of which is connected to the end of the piston shaft. {The piston {(DSBC-U-125-2500-PA-N3, FESTO$^\circledR$)} has an inner diameter of 125 mm and a length of 2500 mm}. The outlet of the piston is connected to the inlet tank through a hose. This tank, made of acrylic plates, has a dimension of 340~mm $\times$ 340~mm $\times$ 500~mm, and is taken to decay the flow disturbance in experiments and remove bubbles when filling up the fluid into the setup. {A contraction nozzle (in area ratio of $9:1$) connects the inlet tank with the channel.} The two motors are precisely synchronized to ensure the same Reynolds number in both inlets (and thus in the mixing channel) and their motions were calibrated. {The Reynolds number depends on the bulk velocity (depending on the piston diameter and motor speed), the kinematic viscosity of water (measured by Ubbelohde viscometer) and the height of the T-mixer channel. {Fluid temperature is monitored at both inlet tanks, and the temporally averaged temperature is 294.65~K with a standard deviation of 0.2~K.} According to the propagation of uncertainty, the relative uncertainty of $Re$ in the present measurements is below 3\%. The details of the uncertainty of $Re$ can be found in \citet{Li2023Thesis}}. The maximum dimensionless running time of the pistons is about $400$($H/U_0$), {which is given by the {piston volume}. } 

\subsection{Specification of the PIV measurements}

For the PIV measurements, hollow glass spheres with a median diameter of 11~$\mu$m are used as tracers. A high-repetition pulse laser with a wavelength of 527~nm (Photonics Industries, dual-head Nd:YLF, maximum pulse energy 50~mJ at the repetition rate 1~kHz) is used to generate light sheet illumination through {a group of concave, convex and cylindrical lenses} \citep{raffel2018particle}. A high-speed camera (Phantom VEO 640, 36~GB onboard RAM, maximum frame rate of 1490~Hz at the full resolution $2560 \times 1600$ pixel$^2$), equipped with a macro-lens ($f=100$~mm), is used for the recording of tracer images. {A bandpass filter (center wavelength 530 $\pm$ 10 nm) is mounted in front of the camera lens to block the room light signal.} An imaged tracer is about 3~pixels in size to avoid the peak-locking issue in the displacement \citep{raffel2018particle}. The synchronization of the laser and camera is precisely controlled using a programmable timing unit from LaVision. The image sampling rate was adjusted for each flow Reynolds number to ensure that the maximum tracer displacement ranged from 6 to 8 pixels. The PIV vector fields are obtained through a multi-step algorithm with the final interrogation window of $32^2$~pixels$^2$ (50$\%$ window overlap) using LaVision Davis software, {where the measurement uncertainty is expected to be around 0.1 pixel and smaller \citep{Wieneke2015}. The magnification is determined by the height/width of the channel. The PIV measurements at the junction were carried out at the $y-z$ plane ($x/H = 0$) and at both inlets ($y-z$ plane, the measurement field-of-view centered at $y/H = \pm 2.5$). The measurements at $x/H = 8$, $12$, $16$ in the outlet channel were carried out in $x-y$ plane ($z/H = 0$).}
Table~\ref{tab:exp_config} summarizes the experiments conducted in this work.

\begin{table}[]
	\centering
	\begin{tabular}{ccccc}
		\hline
		%\multirow{2}{*}{$Re$} & \multirow{2}{*}{$H$} & \multirow{2}{*}{$U_0$} & \multirow{2}{*}{Sampling frequence} & \multirow{2}{*}{Cumulative trials} & \multicolumn{2}{c}{Measurement time of one trail} \\ %\cline{6-7} 
		%&  &  &  &  & At inlet channels & At junction and outlet channels \\ 
		%- & {[}mm{]} & {[}mm/s{]} & {[}Hz{]} & - & {[}$D/U_0${]} & {[}$D/U_0${]} \\
		$Re$ {[}-{]}  & $U_0$  {[}mm/s{]} &  sampling rate {[}Hz{]} & time/run  {{[}$H/U_0${]}} & time/run {{[}$H/U_0${]}} \\
		&&&inlet&outlet\\
		\hline
		100  & 2.5 & 25  & - & 15.5 \\
		160  & 4.0 & 25  & - & 14.0 \\
		200  & 5.0 & 25  & - & 14.0 \\
		220  & 5.5 & 25  & - & 15.4 \\
		240  & 6.0 & 25  & - & 16.8 \\
		260  & 6.5 & 25  & - & 18.2 \\
		280  & 7.0 & 50  & - & 14.0 \\
		300  & 7.5 & 50  & 50.0 & 15.0 \\
		320  & 8.0 & 50  & - & 16.0 \\
		340  & 8.5 & 50  & - & 17.0 \\
		360  & 9.0 & 50  & - & 18.0 \\
		380  & 9.5 & 50  & - & 19.0 \\
		500  & 12.5 & 75  & 50.0 & - \\
		700  & 17.5 & 75  & 50.0 & - \\
		900  & 22.5 & 100  & 50.0 & - \\
		1100  & 27.5 & 100  & 50.0 & - \\
		1500  & 37.5 & 100  & 58.0 & 58.0 \\
		2000  & 50.0 & 300  & 25.8 & 25.8 \\
		3000  & 75.0 & 400 & 29.0 & 29.0 \\
		4000  & 100.0 & 500  & 31.0 & 31.0 \\
		5000  & 125.0 & 600  & 32.0 & 32.0 \\ \hline
	\end{tabular}
	\caption{PIV measurements conducted in this work. The velocity field was measured at the inlet and/or outlet. For each case, two runs were performed to get well-converged statistics. In addition, PLIF measurements were conducted for $1500\le Re \le 5000$ with the same specifications as for the PIV measurements for the outlet case. }
	\label{tab:exp_config}
\end{table}

\subsection{Specification of the PLIF measurements}\label{Sec:PLIF}

PLIF was applied to measure the scalar concentration field {at downstream positions $x/H = 8$, 12, 16 in the outlet channel}. For this purpose, one inlet branch (including the piston and the tank) was filled with dilute fluorescent dye ({60 $\mu$g/L}), Rhodamine 6G, with Schmidt number of $600-1250$ \citep{Odier2014JFM, Crimaldi2001EF}. The same high-repetition rate pulse laser and camera as in the PIV measurements were used for the PLIF measurements.  The camera was equipped with {a long-pass filter (with a 550 nm cutoff}) to reduce the background light. The depth resolution was given by the thickness of the laser sheet. To approach the Batchelor scale in this direction, we used a Gaussian-shape neutral density (ND) filter ({NDYR20B, Thorlabs}) in the path of {the collimated beam of the laser} to obtain a sharper Gaussian profile of the laser beam, before expanding it {with a spherical convex lens and a cylindircal concave lens} into a light sheet. For a top-hat profile beam, this filter provided a Gaussian profile beam. For an ideal Gaussian profile beam, this neutral density filter gave a squared Gaussian profile for the beam. In practice, this filter is expected to give a profile in-between these two cases.
Our configuration results in approximately $0.02$~mm thickness at the waist line from the conservative estimation \citep{ready1997industrial, crimaldi2008planar, Lavertu2008JFM}. {The resulting Rayleigh length is about 2~{mm}, and for the area within Rayleigh length the averaged thickness of the laser sheet (about $0.025$~{mm}) is close to the Batchelor scale ($\eta_b\approx 0.02$~{mm}) at $Re=650$, estimated conservatively using the peak dissipation of the turbulent kinetic energy from direct numerical simulations of the T-mixer flow in the same configuration \citep{schikarski2017direct, schikarski2019inflow}. The Rayleigh area has dimensions of about $24$~{mm} ($x$) $\times$ $2$~{mm} ($y$) and is centered at $y/H=0$ in the PLIF measurements for the outlet channel flow.}
%\textcolor{red}{State the estimated Batchelor scale specifically; it is also unclear what dimensions the Rayleigh area has. Neither it is stated where this area is centered at. This would imply that the measurement has different precision at different locations, which is OK, because the camera resolution is the limiting factor, right? I think it is worth commenting on all this.} 
To better observe the macroscopic mixing behaviours at high $Re$ for the same dimension of field-of-view as the PIV measurement, {the PLIF camera was placed approximate 0.5 m away from the laser sheet plane and gave the image pixel resolution about 1.6 times the Batchelor scale (0.05 times the Kolmogorov scale) for $Re\leqslant650$}. For the $Re=5000$, the largest Reynolds number in this study, the pixel resolution was estimated to be about $11.5\eta_b$. 

\begin{figure*}
	\centering
	\includegraphics[width=0.45\textwidth]{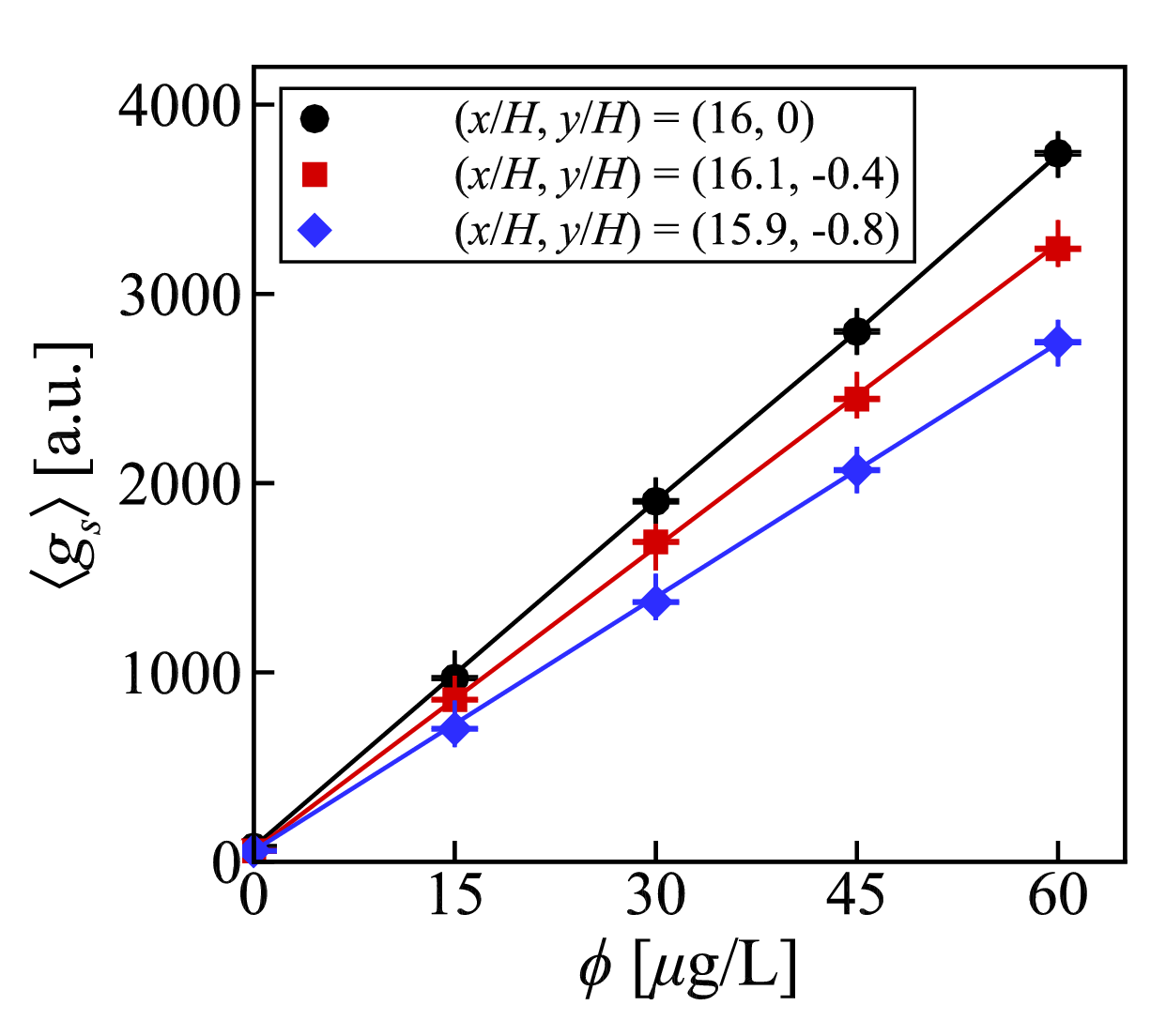}
	\caption{\RF{(Color online) Characteristic PLIF calibration curves at three points as examples. Symbols denote the temporally averaged grayscale values from measurements, with vertical error bars indicating the root-mean-square values. Horizontal error bars indicate the uncertainty in preset concentrations. Solid lines indicate the corresponding linear fits.}}\label{fig:calib}
\end{figure*}
For high Reynolds number experiments $Re\ge 1500$, {to enable the high-frequency, long-duration PIV and PLIF measurements for turbulent statistics, a portion of the camera's full field-of-view (2480 $\times$ 500 pixel$^2$) was employed to capture the flow and mixing in the T-mixer outlet channel. The results are presented in \S\ref{sec:turb} and \S\ref{sec:sca}, respectively. For the PLIF calibration, the whole T-mixer channels were filled with well dissolved dye solution in a preset concentration, and the camera is used to record PLIF images. \RF{The preset concentrations were prepared using graduated cylinders and an electronic balance. The dye concentration ranges from $0$ to $60$~$\mu${g/L} with a $15$~$\mu${g/L} increment. At each concentration, $1000$ snapshots were recorded, and a linear relationship was established between the temporally averaged grayscale value $\langle g_s \rangle(m,n)$ in the PLIF images and the preset concentration $\phi$ for each pixel, i.e., $\langle g_s\rangle(m,n) = \varGamma(m,n) \phi + \langle g_b \rangle(m,n)$. Here, $\varGamma(m,n)$ denotes the system-specific optical configuration and $\langle g_b \rangle$ is the temporally averaged grayscale value corresponding to the camera background noise, and they are obtained from the least-squares linear fitting for each pixel, given by the approximately linear relation as shown in figure~\ref{fig:calib}. 
In the mixing measurements, the hardware configurations are kept unchanged from the calibration, and $\varGamma(m,n)$ and $\langle g_b \rangle(m,n)$ are assumed to be unchanged. The instantaneous concentration field can be obtained as 
\begin{equation}\label{eq:concentration}
	\phi(m,n,t) = [g_s(m,n,t)-\langle g_b \rangle(m,n)]/\varGamma(m,n).
\end{equation}
The uncertainty of the instantaneous concentration field $\sigma_\phi$ for each pixel can be evaluated with the uncertainty of each term in equation~\ref{eq:concentration} in the propagation of uncertainty \citep{Tavoularis2005},
\begin{equation}\label{eq:calib_error}
	\begin{aligned}
		\sigma_\phi = & \bigg[\bigg(\frac{\partial \phi}{\partial g_s}\sigma_{g_s}\bigg)^2 + \bigg(\frac{\partial \phi}{\partial \langle g_b \rangle}\sigma_{\langle g_b \rangle}\bigg)^2 + \bigg(\frac{\partial \phi}{\partial \varGamma}\sigma_{\varGamma}\bigg)^2 \bigg]^{1/2} \\
		=& \bigg[\bigg(\frac{1}{\varGamma}\sigma_{g_s}\bigg)^2 + \bigg(\frac{1}{\varGamma}\sigma_{\langle g_b \rangle}\bigg)^2 + \bigg(\frac{\phi}{\varGamma}\sigma_{\varGamma}\bigg)^2 \bigg]^{1/2},
	\end{aligned}
\end{equation}
where $\sigma_{g_s}$, $\sigma_{\langle g_b \rangle}$ and $\sigma_{\varGamma}$ denote the uncertainties of the measured grayscale value, the background grayscale value and the calibration coefficient, respectively. Given that the hardware configuration of the mixing measurement remains the same as the calibration, the uncertainty source and level for each term is assumed to be same. The root-mean-squares  of $g_s$ and $\langle g_b \rangle$ are taken as $\sigma_{g_s}(=\sigma_{\langle g_s \rangle})$ and $\sigma_{\langle g_b \rangle}$ and the $95\%$ confidence interval of the linear fitting for $\varGamma$ in the calibration is taken for $\sigma_{\varGamma}$. The uncertainty analysis was performed for all images, giving the relative uncertainty ($\sigma_\phi/\phi(m,n,t)$) of the concentration measurements up to about 1.2\% (see method details in \citet{xu2012experimental}). }
 
\subsection{Direct numerical simulation}

A direct numerical simulation of the T-mixer flow was performed at $Re = 1500$ for cross-validation purposes. The numerical configuration was exactly as in the experiments, except for the lengths of the inlet and outlets ducts. At the inlets, the fully developed laminar profile was enforced as boundary condition, which enables to take an inlet length of $3H$ without affecting the flow at the junction \citep{schikarski2017direct}. The outlet was taken to be $17H$ long and a stabilized outflow condition was imposed \citep{DONG}. The dimensionless Navier–Stokes equations were solved with the high-order open-source computational fluid dynamics code Nek5000 \citep{Fischer2005}. The geometry was discretized with 148,087 elements and a polynomial order of 7 was chosen to ensure that the turbulent flow in the outlet channel was well resolved. An adaptive time-stepping approach was used to bound the Courant--Friedrichs--Lewy ($CFL<0.4$) during the simulation. The simulation data were stored at each time step over a period of $450H/U_0$ after an initial transient phase of $50H/U_0$.

\section{Measurements of the velocity field}\label{sec:vel}

\citet{schikarski2019inflow} demonstrated the importance of the inflow conditions in determining the mixing efficiency in the turbulent regime of the T-mixer. Accordingly, we begin by reporting PIV measurements of the inflow conditions at the inlet channels in~\S\ref{sec:inlet}.  Subsequently, we present PIV measurements of the flow patterns in the low Reynolds number regime in~\S\ref{sec:lowRe}, thus validating our experiments against previous works. The turbulent regime is analyzed in detail in \S\ref{sec:turb}.

\subsection{Inflow conditions}\label{sec:inlet}

\begin{figure*}
	\centering
	\includegraphics[trim=0.1cm 0.2cm 0.5cm 0.2cm, clip, width=1\textwidth]{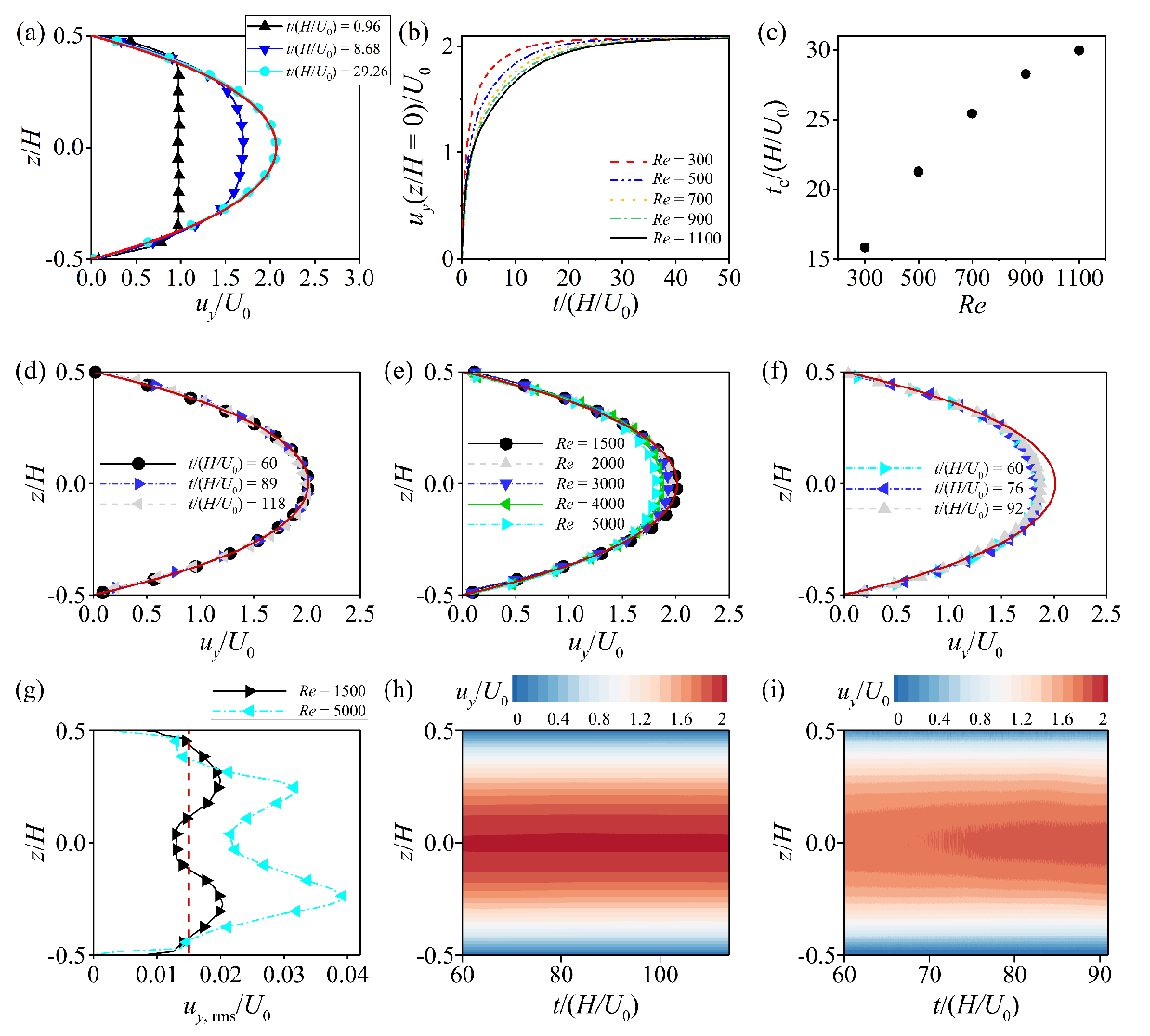}
	\caption{(Color online) (a) {Velocity profiles obtained by PIV measurements at $Re = 700$ and $y/H = 2.5$, where the red line corresponds to the analytical laminar solution} from \citet{shah2014laminar}. (b) Time series of streamwise velocity at $(y/H, z/H)=(2.5,0)$ and $Re = 700$. (c) Critical time for temporally fully developed velocity against the Reynolds number, where the time corresponding to $99\%$ of the final steady velocity in (b) is taken as the critical time $t_c$. (d) Velocity profiles at $Re = 1500$. (e) Velocity profiles at $y/H = 2.5$ for $Re = 1500-5000$ and $t/(H/U_0) = 60$, where the red line represents the same profile as in (a). (f) Velocity profiles at $Re = 5000$. (g) Profiles of the root-mean-square streamwise velocity at $y/H = 2.5$. The red dashed line indicates the mean uncertainty of the PIV measurement. Space-time velocity contours of the streamwise velocity at $y/H = 2.5$ for $Re = 1500$ (h) and $Re = 5000$ (i).}\label{fig:profile}
\end{figure*}

In what follows, we examine the flow {in the plane ($x/H = 0$) of inlet channels} near the junction ($y/H=\pm2.5$) with PIV measurements. At $t/(H/U_0)=0$, the pistons were accelerated rapidly to a preset speed, and the flow in the inlets experienced a transient process to approach a fully developed velocity profile, see figure~\ref{fig:profile}(a). The duration of the transient process increases with the Reynolds number, as shown in figures~\ref{fig:profile}(b--c). As shown in figure~\ref{fig:profile}(d), the fully developed velocity profile measured at $Re=1500$ agrees very well {with the analytical laminar solution \citep{shah2014laminar}, indicated by the red line}. The measurement was started at $t/(H/U_0)=60$ and did not exhibit any temporal fluctuations till $t/(H/U_0)=118$, at which point the camera on-board RAM was full and the measurement stopped. In similar measurements for increasing $Re$, we observed that the development length was not sufficient and the velocity gradually deviated from the analytical solution, see figure~\ref{fig:profile}(e). However, even at $Re=5000$, $u_y(z/H=0)$ reached about $90\%$ of the analytical solution, see figure~\ref{fig:profile}(f). At first sight, this is surprising given that turbulence can already be triggered in square ducts at about $Re\sim 1400$ in the form of patches, if sufficiently perturbed {and having a sufficient development length}, and reach a fully turbulent state at $Re=3000$ \citep{barkley2015rise}. However, the laminar flow in a square duct is linearly stable at all $Re$ \citep{tatsumi1990stability}. In addition, less developed (flatter profiles) tend to be more stable, e.g., in inlet tubes laminar flow was reported up to $Re=100,000$ in experiments \citep{pfenninger1961boundary}.  

In figure~\ref{fig:profile}(g), we show the profiles of the root-mean-square streamwise velocity ($u_{y,\mathrm{rms}}$) for $Re = 1500$ and $5000$. At $Re=1500$, the $u_{y,\mathrm{rms}}$ is small, close to the PIV uncertainty, whereas for $Re=5000$ the fluctuations are non-negligible. A spatio-temporal representation of the streamwise velocity profile is shown in figures~\ref{fig:profile}(h) and (i), for $Re = 1500$ and $Re = 5000$, respectively. They confirm that the flow had negligible fluctuations at $Re=1500$, whereas at $Re=5000$ it appears that the flow was still slightly developing, while exhibiting incipient signs of transition.  We interpret these results as follow. Residual swirling motions from the hose in the tank (figure~\ref{fig:sketch}b) may initially decay along the inlet and then begin to grow as the profile develops. However, turbulence cannot develop before entering the junction, because of the 75$H$-long inlets of our setup. For example, \citet{barkley2015rise} perturbed the flow 120$H$ after the entrance and observed fully developed turbulence further 100$H$ downstream. Without disturbance their duct remained laminar up to $Re=5000$. In summary, for all cases in this study, the measurements begin from $t=60H/U_0$, and the inlet flows before entering the T-junction can be considered quasi-laminar (developed up to 90\% and with 4\% fluctuation at the highest  Reynolds number investigated, $Re=5000$). 

\subsection{Characteristic flow patterns of the low Reynolds number regime}\label{sec:lowRe}

We validated our setup with PIV measurements at low Reynolds numbers, $Re<400$. The cross-sectional  velocity field, $(u_y,u_z)$, was measured in the $y-z$ plane at $x/H=0$, and the recording lasted for 14$H/U_0$. For $Re = 100$, the flow velocity field was found to be approximately left-right and top-down symmetric, as shown in figure~\ref{fig:regime}(a), whereas for $Re=160$ ({shown in figure~\ref{fig:regime}b}) both reflection symmetries are broken, while the flow remains symmetric to a $180^\circ$ rotation, corresponding to the engulfment regime. Within the measurement duration, the flow remained stationary. Note that due to the symmetry-breaking of the underlying Pitchfork bifurcation, two realizations of the engulfment pattern are possible: the one shown in figure~\ref{fig:regime}(b), and its mirror-symmetric. In agreement with the observations in \citet{thomas2010experimental}, both patterns were observed in our experiments, indicating that our setup does not have a substantial imperfection causing one of the two states to be strongly preferred. At $Re=280$, the flow also featured a similar engulfment pattern with the same rotational symmetry, shown in figure~\ref{fig:regime}(c), however it oscillated periodically in time. At $Re=360$, the flow regained the reflection symmetries (see figure~\ref{fig:regime}d) and was periodic in time. 

\begin{figure*}
	\centering
	\includegraphics[width=0.8\textwidth]{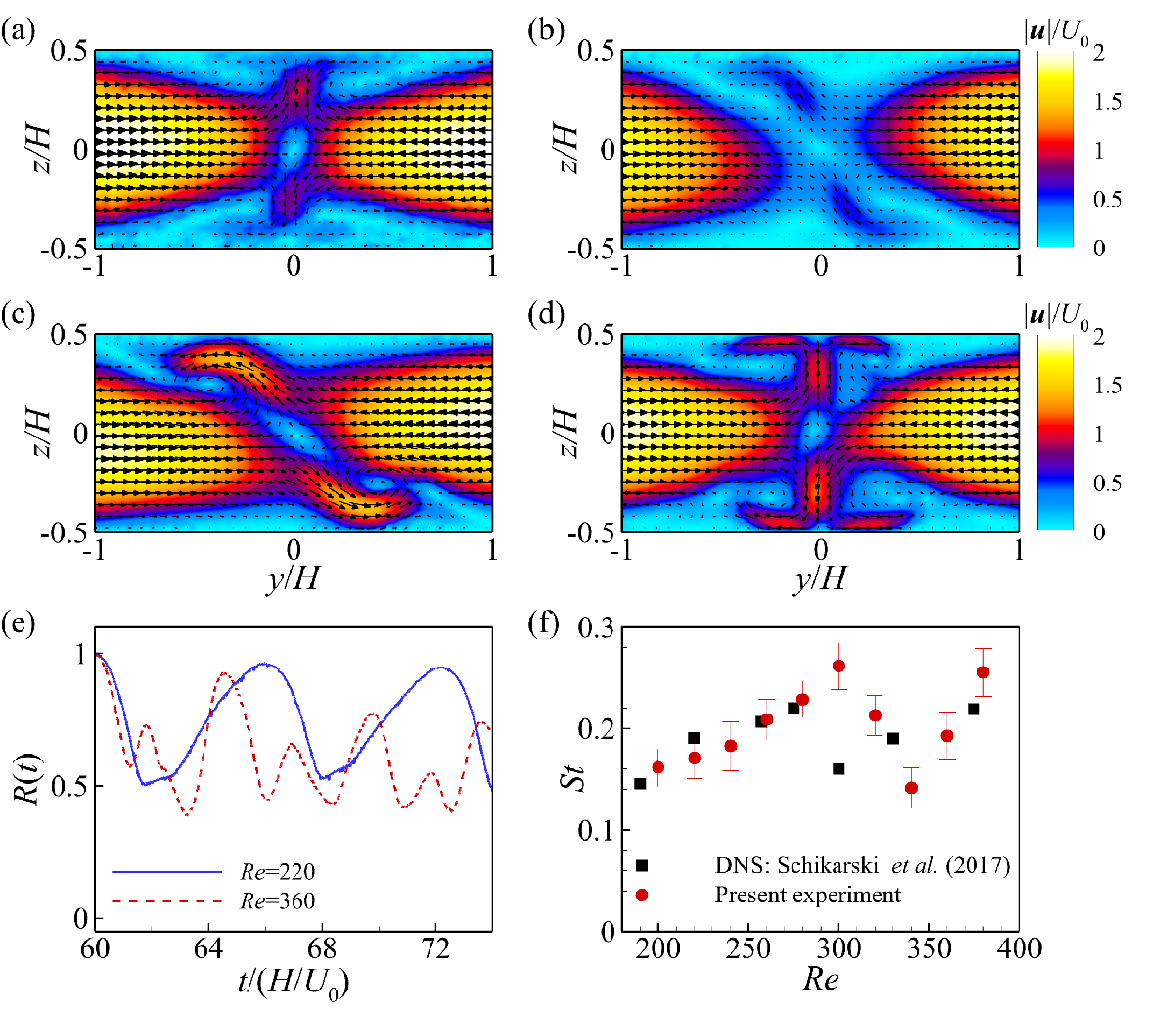}
	\caption{(Color online) Velocity field at the center of the junction for $Re = 100$ (a), $Re = 160$ (b),  $Re = 280$ (c) and $Re = 360$ (d), where the contours show the magnitude of the velocity and (a--d) share the same colormap. (e) The correlation coefficient $R(t)$ against time. (f) Strouhal number corresponding to the periodicity of $R(t)$ {depending on} the Reynolds number.}\label{fig:regime}
\end{figure*}

The reported scenario is in very good qualitative agreement with  previous experimental measurements and direct numerical simulations (DNS) for the same geometry \cite{schikarski2017direct, zhang2019investigation}. To provide a quantitative comparison, the periods of the unsteady flow patterns were extracted. Specifically, each velocity field $|\bm{u}|(m,n,t)$ in the time series was correlated to that of the first snapshot as
\begin{equation}
	R(t)=1-\frac{1}{M \cdot N} \sum_{m=1}^{M} \sum_{n=1}^{N}\left[|\bm{u}|(m, n, t)-|\bm{u}|(m, n, t/(H/U_0)=60)\right]^{2}/U_0^2,
\end{equation}
where $m$ and $n$ are the coordinates of the PIV vector grids, $M$ and $N$ the number of the vectors along each dimension, respectively \citep{thomas2010experimental}. As shown in figure~\ref{fig:regime}(e), either the curve of $Re=220$ or that of $Re=360$ shows distinct periodicity, corresponding to the observation in the animation of the flow velocity field. The fast Fourier transform operation was applied to $R(t)$ and the dominant frequency ${\omega}$ was extracted. Correspondingly, the dimensionless Strouhal number $St = {\omega} H/U_0$ is obtained. The $St$ increases with the increase of $Re$ until $Re=300$, then decreases followed by an increase when $Re$ is further increased. The $St$ trend in our experiments agrees well with the DNS study of \citet{schikarski2017direct}, as shown in figure~\ref{fig:regime}(f). Note that in the range $Re\in[300,400],$ there is hysteresis and both the unsteady engulfment and symmetric regimes can be realized depending on the initial conditions \citep{schikarski2017direct}, or even a mixed state in which the flow randomly jumps between the two, if the inlet conditions do not attain a fully developed laminar flow \citep{schikarski2019inflow}. This regime was not investigated here in detail, as our focus is on high Reynolds numbers, where no hysteresis occurs and the flow is left-right and top-down {symmetric} in average.

We note that in our upscaled setup, there is a small temperature difference between the top and bottom wall of the setup, from the non-uniform temperature along the gravitational direction in the laboratory. This small temperature difference and imperfections in the setup may slightly influence the flow transition among regimes. However, our PIV measurements shown in figure~\ref{fig:regime}(a--d) indicate that such influence is small and it is expected to be even less significant in the turbulent regime. 

\subsection{Analysis of the turbulent velocity field}\label{sec:turb}

\citet{schikarski2019inflow} reported that for $Re>650$ the flow in the outlet channel becomes turbulent. Here, we carried out planar PIV measurements in the outlet channel in the $x-y$ plane ($z/H=0$) centered at the downstream positions $x/H=8, 12, 16$ for $1500\le Re \le 5000$. In figure~\ref{fig:tkemeancontour}, we show colormaps and vector fields of the measured fluid velocity $(u_x,u_y)$ at these three locations for $Re=1500$ and $5000$. To collect the flow statistics presented in this section, two experimental runs were conducted for each $Re$, covering about $64H/U_0$--$116H/U_0$ for each $Re$, depending on the corresponding sampling rates, see table~\ref{tab:exp_config}.

\begin{figure*}
	\centering
	\includegraphics[trim=0.5cm 0.01cm 0.5cm 0.6cm, clip, width=1\textwidth]{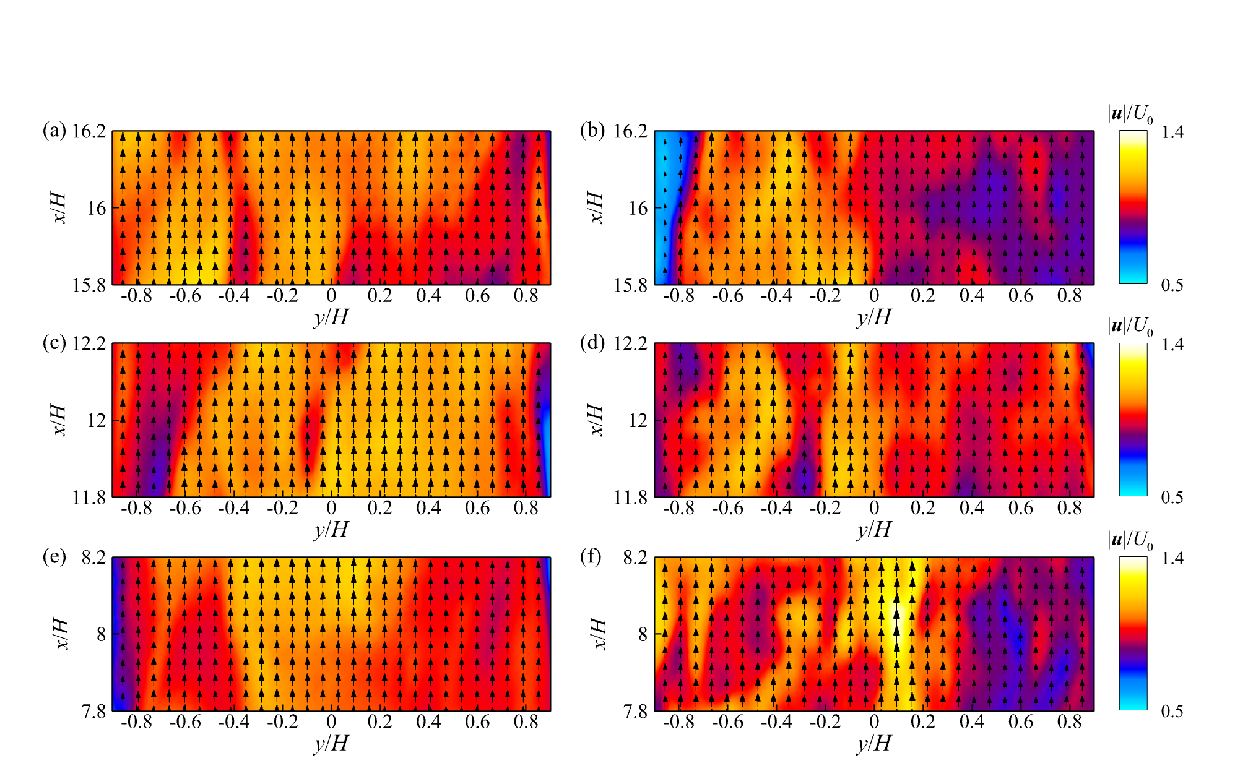}
	\caption{(Color online) Instantaneous snapshots of the velocity field measured at the plane $z/H=0$ for $Re=1500$ (left column) and $Re=5000$ (right column). The measurement planes are centered around three streamwise locations,  $x/H = 16$ (top row), $x/H = 12$ (middle row) and $x/H = 8$ (bottom row). The arrows appear nearly vertical because the instantaneous streamwise velocity component overwhelms the spanwise one ($u_y/u_x<0.08$).}\label{fig:tkemeancontour}
\end{figure*}

%\subsubsection{Mean velocity field}
\begin{figure*}
	\centering
	\includegraphics[trim=0.1cm 0.1cm 0.2cm 0.2cm, clip, width=1\textwidth]{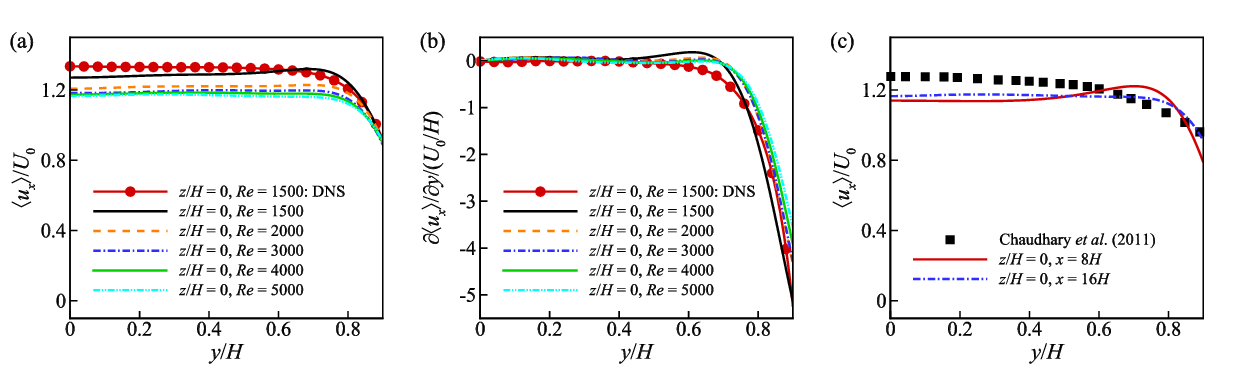}
	\caption{(Color online) Average streamwise velocity profiles $\left<u_x\right>$ (a) and their corresponding lateral gradients (b) for several Reynolds numbers at $x/H = 16$, and as a function of downstream position for $Re = 5000$ (c). The black square symbols represent fully developed turbulent duct flow (of aspect ratio two) at $Re = 5300$ \citep{chaudhary2011direct}. }\label{fig:meanuv}
\end{figure*}
Figure~\ref{fig:meanuv}(a) displays the measured average streamwise velocity $\left<u_x\right>$ for all Reynolds numbers at $x/H = 16$ and $z/H=0$; the average lateral velocity $\left<u_y\right>$ is nearly zero, as expected from the top-down symmetry of the turbulent state, and is not shown. As $Re$ increases, the velocity profile gradually flattens and the gradient near the wall sharpens, as evidenced in figure~\ref{fig:meanuv}(b). Our DNS at $Re=1500$ is in good qualitative agreement with the experiment, however, it decreases monotonously toward the wall, whereas the experiment features a maximum at about $y/H=0.7$. {This slight discrepancy may be attributed to differences in the boundary conditions and the setup configuration between the DNS and the experiment.} As shown in figure~\ref{fig:meanuv}(c), the velocity profiles measured in the experiment clearly deviate from the fully developed velocity profile in turbulent duct flow \cite{chaudhary2011direct} {(possibly due to the insufficient developed length)}. Thus, turbulence in the outlet channel is still far from equilibrium at $x/H=16$ yet.

%\textcolor{red}{I'm not sure the following text is correct: The DNS at 1500 does not show this jump, Mehdi could you plot profiles at earlier cross sections as done for the experiments at $Re=1500$?	Figure~\ref{fig:meanuv}(c) shows $\langle u_y \rangle$ at $y/H=8$, $12$ and $16$ for $Re = 5000$. At $y/H=8$, $\langle u_y \rangle$ shows a bump around $x/H\approx0.7$, which differs to the nearly collapsed profiles at $y/H=12$ and $16$. A similar bump can be also observed at $y/H=16$ for $Re=1500$ (see figure~\ref{fig:meanuv}b). This bump results from the streams of the inlet flows. For the downstream locations, the momentum of the stream is re-distributed by the later momentum transportation. The larger Reynolds number, the shorter streamwise distance in the outlet channel is needed to dissipate the influence of the inlet streams. Once the signature of the inlet streams disappears in the velocity profile, the influence of inlet condition on the flow statistics is expected to be trivial at further downstream locations. }

%\subsubsection{Turbulent kinetic energy and dissipation}\label{sec:TC}
\begin{figure*}
	\centering
	\includegraphics[trim=0.1cm 1.3cm 0.2cm 0.2cm, clip, width=1\textwidth]{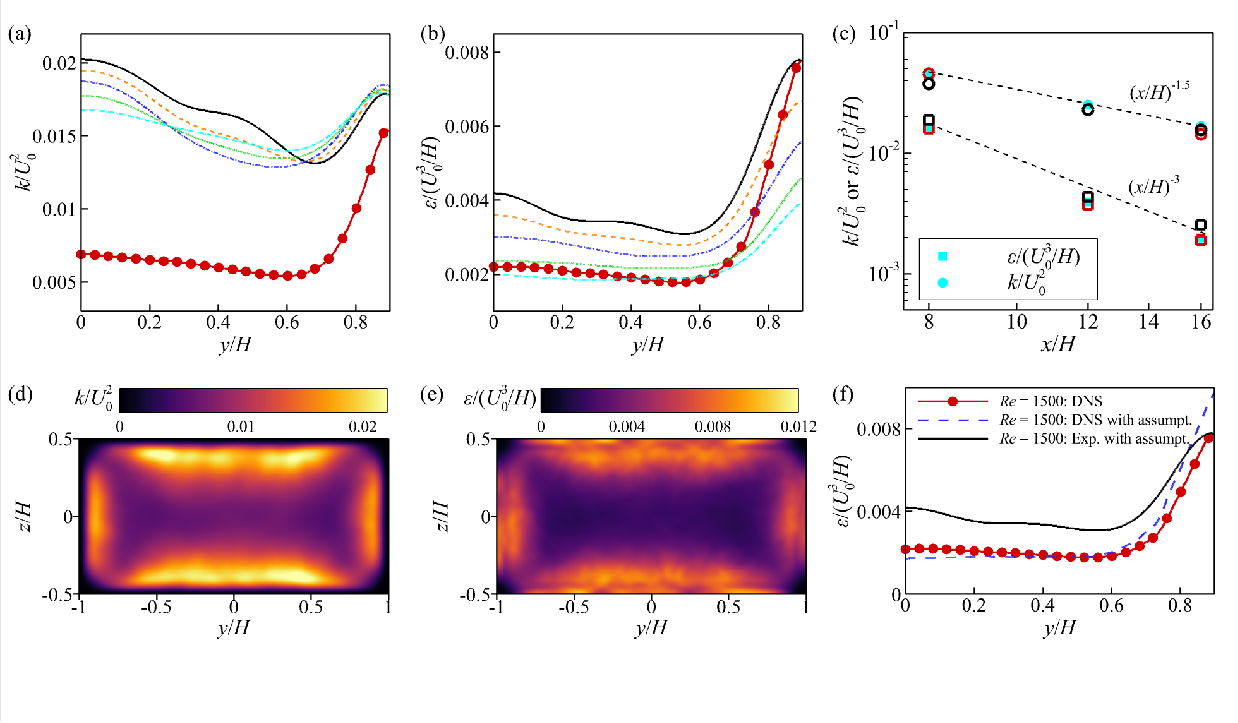}
	\caption{(Color online) (a) Profiles of turbulent kinetic energy ($k$) at $x/H = 16$ and several $Re$, as indicated in the legend. (b) Profiles of the (estimated) turbulent dissipation rate ($\epsilon$) at $x/H = 16$. {The legends of (a) and (b) match those in figure~\ref{fig:meanuv}(a).} (c) Decay of turbulent kinetic energy (circle symbols) and turbulent dissipation rate (square symbols) along the streamwise direction at $y/H = 0$ (cyan), $y/H = 0.5$ (red hollow) and $y/H = 0.75$ (black hollow) in the outlet channel for $Re = 5000$. (d)-(e) Colormaps of the turbulent kinetic energy and dissipation at $x/H=16$ from DNS at $Re=1500$. (f) The comparison of the dissipation profiles computed with the usual definition and with the assumption in table~\ref{tab:ass}.}\label{fig:tkeudiss}
\end{figure*}

The cross-sectional turbulent kinetic energy (TKE) $k$ was calculated using the measured root-mean-square velocity fluctuations, $k = \left(\left<u_x^{\prime 2}\right> + \left<u_y^{\prime 2}\right> \right)/2$. Given that $k$ varies weakly along the streamwise ($x$) direction within our measurement plane (of length $0.4H$), $k$ was averaged along $x$ {and determined at $z/H = 0$}. The so-computed spanwise profiles of $k$ are shown in figure~\ref{fig:tkeudiss}(a). At $Re=1500$, the TKE is large at the center of the outlet channel and decreases across the channel width, reaching the minimum around $y/H \in[0.6,0.7]$, then the TKE increases as the wall is {approached}. This trend holds for all the Reynolds numbers in the experiments. As $Re$ increases, $k(y/H=0)$ decreases, while $k$ close to the side wall is nearly unchanged.

The viscous dissipation of turbulent kinetic energy $\epsilon$ was also examined. \RF{As it is not possible to obtain the out-of-plane velocity component from our PIV measurements, $\epsilon$ was estimated with two different assumptions: the assumptions of local axisymmetry and local isotropy \citep{George&Hussein1991JFM, Doron2001JPO, xu2013}; the corresponding formulas are listed in table~\ref{tab:veldissipation}. The $\epsilon$ from the two estimation methods yield similar results (within about $5\%$ difference), and their average was taken for the analysis;} as for $k$, $\epsilon$ was also averaged along the streamwise direction within the measurement field-of-view. The results are shown in figure~\ref{fig:tkeudiss}(b). At $Re=1500$, the dissipation has a rather flat profile close to the channel center. The dissipation increases rapidly for $y/H>0.7$ and reaches the corresponding maximum close to the side wall, where the shear is strong. As $Re$ increases, the (dimensionless) dissipation gets smaller, in general agreement with the DNS of \citet{schikarski2019inflow} that the cross-sectionally averaged dissipation rate decreases slightly with the increase of the Reynolds number (see their figure 10b). The turbulent kinetic energy and the dissipation at $y/H = 0$, $0.5$, and $0.75$ were extracted at $x/H=8$, $12$ and $16$ for $Re=5000$. As shown in figure~\ref{fig:tkeudiss}(c), the turbulent kinetic energy decays along the streamwise direction, confirming that the turbulence here differs greatly from the equilibrium turbulence in a duct. The decay of turbulent kinetic energy exhibits an approximate $(x/H)^{-1.5}$ scaling. Similarly, the dissipation also decays along the streamwise direction, following an approximate $(x/H)^{-3}$ scaling.

\begin{table}[]
	\centering
	\caption{Assumptions used for estimating the dissipation from PIV data.}
	\label{tab:veldissipation}
	\begin{tabular}{ll}
		\toprule
		Assumption & Expression  \\
		\midrule
		%Isotropy & $
		%\begin{aligned}
		%\epsilon = 15\nu \left<\left(\frac{\partial u^\prime_y}{\partial y}\right)^2\right>
		%\end{aligned}
		%$  \\
		Local axisymmetry & $
		\begin{aligned}
			\epsilon=\nu\left\langle-\left(\frac{\partial u_x^{\prime}}{\partial x}\right)^2+8\left(\frac{\partial u_y^{\prime}}{\partial y}\right)^2+2\left(\frac{\partial u_x^{\prime}}{\partial y}\right)^2+2\left(\frac{\partial u_y^{\prime}}{\partial x}\right)^2\right\rangle
		\end{aligned}
		$  \\
		Local isotropy & $
		\begin{aligned}
			\epsilon= & \nu\left\langle 4\left(\frac{\partial u_x^{\prime}}{\partial x}\right)^2+4\left(\frac{\partial u_y^{\prime}}{\partial y}\right)^2+3\left(\frac{\partial u_x^{\prime}}{\partial y}\right)^2+3\left(\frac{\partial u_y^{\prime}}{\partial x}\right)^2\right. \\
			& \left.+4\left(\frac{\partial u_x^{\prime}}{\partial x} \frac{\partial u_y^{\prime}}{\partial y}\right)+6\left(\frac{\partial u_x^{\prime}}{\partial y} \frac{\partial u_y^{\prime}}{\partial x}\right)\right\rangle
		\end{aligned}
		$ \\
		\bottomrule
	\end{tabular}
	\label{tab:ass}
\end{table}

Colormaps of $k$ and $\epsilon$ at $x/H=16$ obtained from our DNS at $Re=1500$ are shown in figures~\ref{fig:tkeudiss}(d)-(e), respectively. Here $k$ was computed using the three velocity components and $\epsilon$ using the usual definition. It is apparent from the colormaps that the boundary layers are thicker in the vertical (short) direction than in the horizontal (long) one. The line profiles of $k$ and $\epsilon$ at $(x/H,z/H)=(16,0)$ are compared in figure~\ref{fig:tkeudiss}(a)-(b), respectively, with the corresponding experimental measurements. Note that here $k$ is computed from the numerical data using the two components measured in experiments only (i.e., omitting $u_z^{\prime}$). The DNS show a much weaker turbulence activity near the center than in the experiments (by about a factor of 2 in both $k$ and $\epsilon$). Both quantities decrease slightly toward the wall, until $y/H\approx 0.7$, where they both rise to reach values in quantitative agreement with the experiments. 

The assumption for estimating the dissipation from the experiments was examined by leveraging the DNS data, as shown in figure~\ref{fig:tkeudiss}(f), the dissipation curves of the DNS data obtained with the same assumptions as in the experiment and the computed directly agree well. This validates the dissipation estimation method used for the experimental data. The differences between the experiment and the DNS could be associated with the small differences in the flow conditions at the inlet. Although the inlet flow is laminar in both of the simulation and the experiments, in the experiment the inlet flow may contain perturbations from the tank hose and the channel segment connectors. Indeed,  \citet{schikarski2019inflow} showed that differences in the inlet flow condition can have a strong influence on the flow characteristics in the T-mixer. Further discrepancies could be produced by slight misalignment between the segments comprising the mixing channel.

\section{Measurements of the turbulent scalar field}\label{sec:sca}

PLIF measurements were conducted to measure the scalar field in the turbulent regime at the same Reynolds numbers as in the previous subsection and with the same technical specifications (see table~\ref{tab:exp_config}).

\subsection{Mixture fraction and scalar variance}
\begin{figure}
	\centering
	\includegraphics[trim=0.0cm 0.6cm 0.2cm 0.5cm, clip, width=1\textwidth]{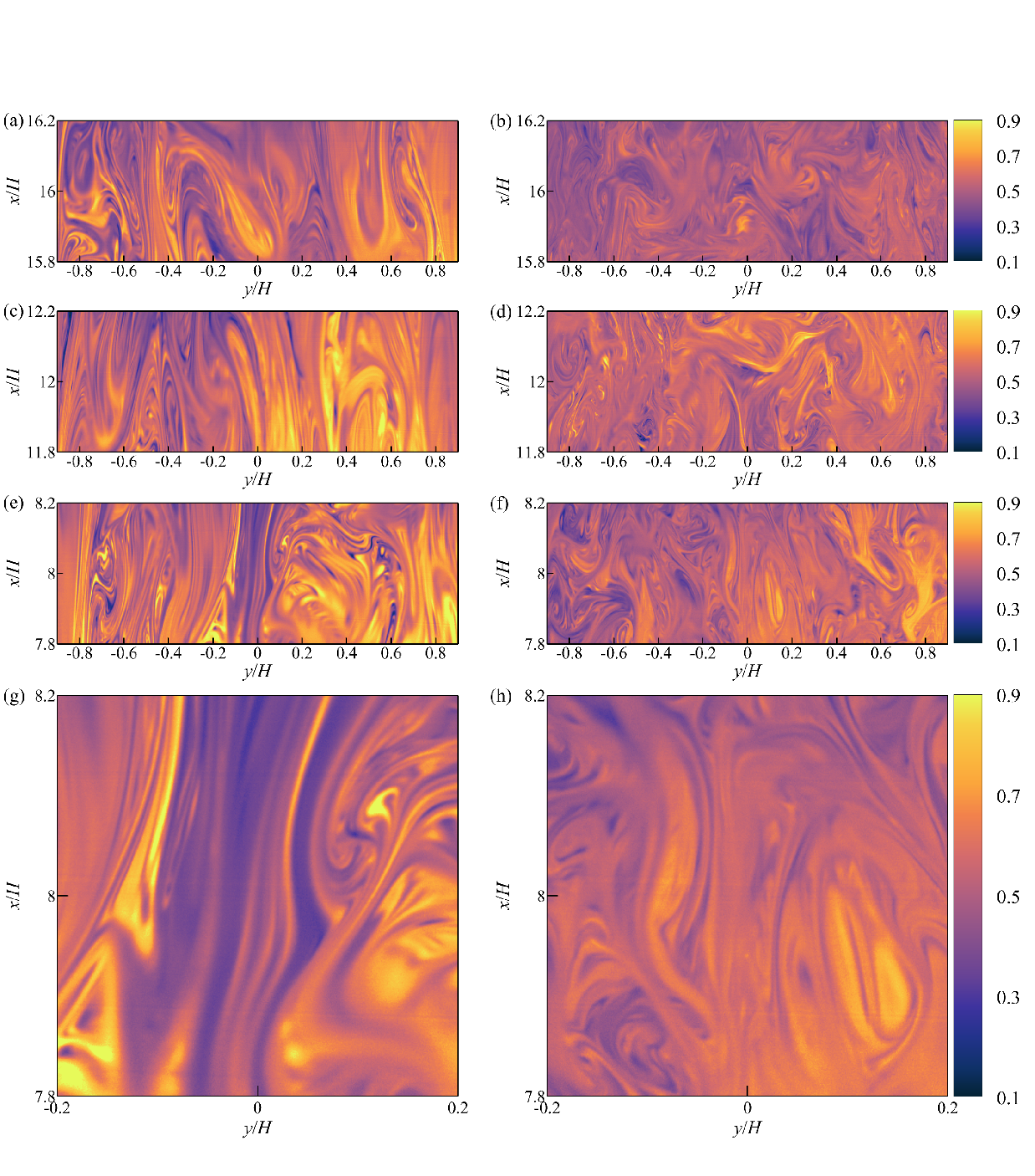}
	\caption{(Color online) Instantaneous snapshots of the scalar field \RF{$\phi/\phi_0$} measured at the plane $z/H=0$ for $Re=1500$ (left column) and $Re=5000$ (right column). The measurement planes are centered around three streamwise locations, $x/H = 16$ (top row), $x/H = 12$ (middle row) and $x/H = 8$ (bottom row), as in figure~\ref{fig:tkemeancontour} where snapshots of the velocity field are shown. {(g, h) The close-up of (e) and (f) around the thinnest of the laser sheet, and the corresponding animations are shown in supplemental movies.}}\label{fig:instan_concentration}
\end{figure}
In figure~\ref{fig:instan_concentration}, we show snapshots of the scalar concentration $\phi$ at $x/H = 16$, $12$ and $8$, respectively. To facilitate comparison, the scalar field is normalized by the initial concentration ($\phi_0$), \RF{correspondingly $\phi/\phi_0=1$ denotes the inlet stream with the fluorescent dye and $\phi/\phi_0=0$ for the stream with pure water}. \RF{When the two streams are fully mixed, $\phi/\phi_0=0.5$}. At $Re = 1500$ (left panel), the scalar field features large-scale structures and the two streams remain partly unmixed, whereas at $Re=5000$ (right) the structure is much finer and the mixing much more complete. Profiles of the {temporally averaged} scalar concentration $\langle\phi\rangle$ at $x/H = 16$ are shown in figure~\ref{fig:MeanMixture}(a). As shown in figure~\ref{fig:MeanMixture}(a), the $\langle \phi \rangle$ profiles exhibit the expected centro-symmetry with respect to $y/H=0$ and become flatter as $Re$ increases. 
The scalar variance, shown in figure~\ref{fig:MeanMixture}(b), is approximately constant along the lateral direction and is reduced by a factor of five, as $Re$ increases from $Re=1500$ to $5000$. This is better seen by plotting its lateral average as a function of $Re$, as shown in the inset of figure~\ref{fig:MeanMixture}(b). Extrapolating our data linearly suggests that the two streams would be fully mixed at $Re=6000$. 

The scalar dissipation \RF{$\epsilon_\phi = 2 D \langle \nabla \phi^\prime \cdot \nabla \phi^\prime \rangle$} at point $(m,n)$ in our measurements was estimated using the finite-difference method of \citet{Buch&Dahm1996JFM}, i.e., 
\RF{
\begin{equation}
	\begin{aligned}
		&\epsilon_{\phi(m, n)} \\ & \approx 2D \Big\langle \Big[\frac{1}{4 \Delta}[\phi^\prime({m+1, n})-\phi^\prime({m-1, n})]+\frac{1}{8 \Delta}[\phi^\prime({m+1, n+1})+\phi^\prime({m+1, n-1})]\\
		&\qquad\; +\frac{1}{8 \Delta}[-\phi^\prime({m-1, n-1})-\phi^\prime({m-1, n+1})] \Big]^2 \Big\rangle \\
		& +2D \Big\langle \Big[\frac{1}{4 \Delta}[\phi^\prime({m, n+1})-\phi^\prime({m, n-1})]+\frac{1}{8 \Delta}[\phi^\prime({m+1, n+1})+\phi^\prime({m-1, n+1})]\\
		& \qquad\;+\frac{1}{8 \Delta}[-\phi^\prime({m-1, n-1})-\phi^\prime({m+1, n-1})]\Big]^2 \Big\rangle,
	\end{aligned}
\end{equation}}
where $\Delta$ is the centre-to-centre pixel spacing in the flow. This eight-point derivative method is equivalent to an implicit filter that reduces the effects of noise on the gradient vector computations \citep{Buch&Dahm1996JFM}. Aside from the implicit noise reduction inherent in this derivative, no explicit smoothing or filtering was applied to the resulting gradient vector fields. The scalar dissipation profile is shown in figure~\ref{fig:MeanMixture}(c). For $Re=1500$, $\epsilon_{\phi}$ has minimum around the channel center and has the maximum close to the side wall. This trend prevails for all Reynolds numbers, and the larger $Re$, the flatter the profile is. In addition, as the Reynolds number increases, $\epsilon_{\phi}$ generally decreases. For $Re=5000$, ${\langle\phi^{\prime}\phi^{\prime}\rangle}$ and $\epsilon_{\phi}$  at $y/H=0$, $0.5$ and $0.75$ are shown at $x/H=8$, $12$ and $16$. ${\langle\phi^{\prime}\phi^{\prime}\rangle}$ and $\epsilon_{\phi}$ decay along $x/H$, following the approximate scalings $(x/H)^{-1.2}$ and $(x/H)^{-0.5}$, respectively. 
\begin{figure*}
	\centering
	\includegraphics[trim=0.0cm 0.0cm 0.0cm 0.5cm, clip, width=0.8\textwidth]{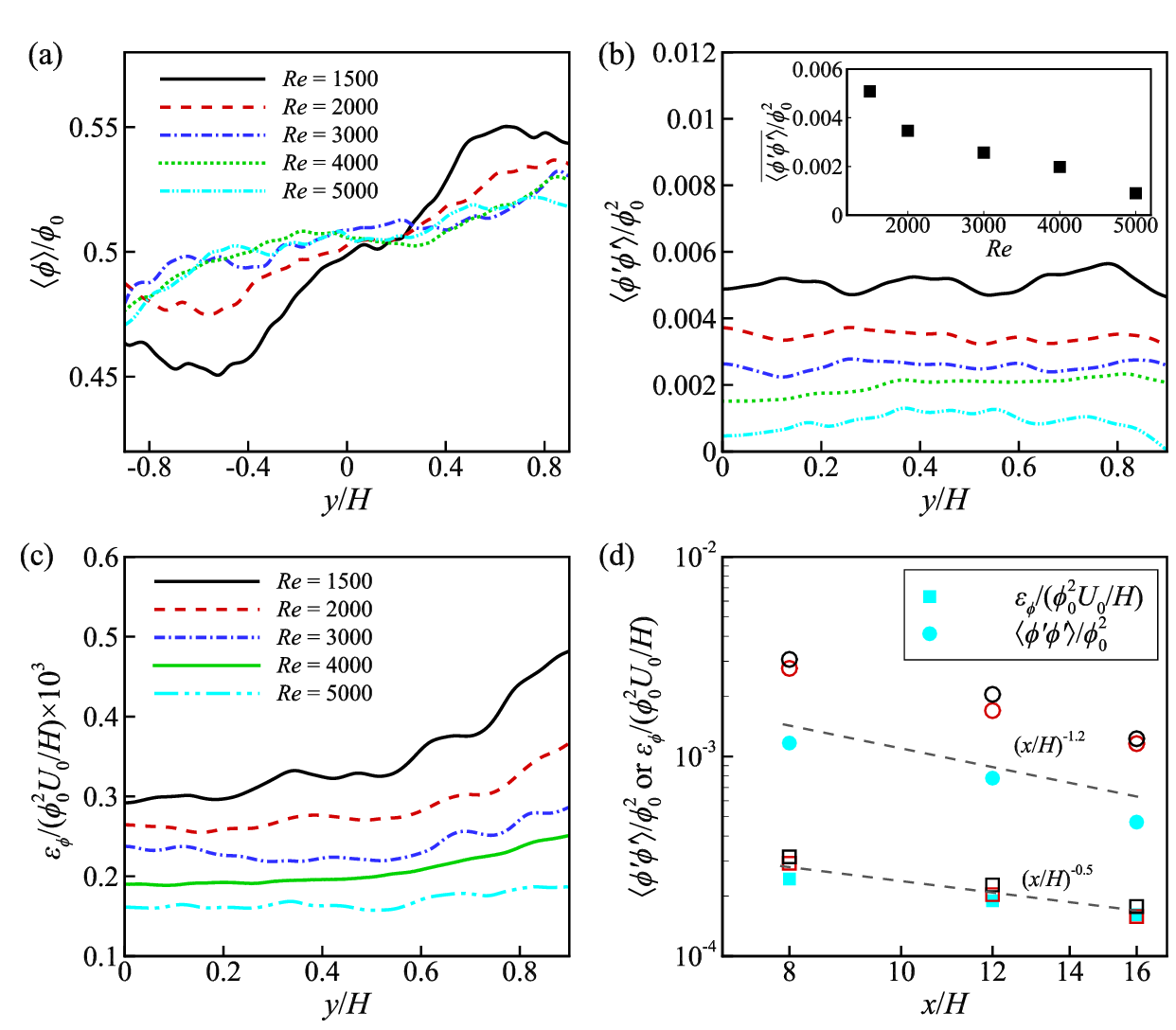}
	\caption{(Color online) The profile of mean mixture fraction $\left<\phi\right>$ (a) and scalar variance $\langle\phi^{\prime}\phi^{\prime}\rangle$ (b) at {$(x/H, z/H) = (16, 0)$}. The inset in (b) shows the mean scalar variance averaged along the $x$ direction as a function of $Re$. (c) The profile of scalar dissipation. (d) Decay of scalar variance (circle symbols) and scalar dissipation (square symbols) at $y/H = 0.0$ (cyan), $y/H = 0.5$ (red hollow) and $y/H = 0.75$ (black hollow) along the outlet channel for $Re = 5000$.}\label{fig:MeanMixture}
\end{figure*}

\subsection{Probability distribution function of the scalar field}\label{sec:onepoint}

The prediction of mixing-sensitive reactions in chemical engineering requires knowledge of the full probability density function (PDF) of the concentration, which allows closing the chemical reaction term in CFD simulations, provided that equilibrium in each fluid particle can be assumed \cite{Pope_2000}. However, in applications detailed measurements of the PDF are usually unavailable and the PDF must be approximated from the scalar mean and variance. Among the various presumed PDFs used for binary mixing, the $\beta$-PDF
\RF{\begin{equation}
	f(\Phi)=\frac{\Phi^{\alpha-1}(1-\Phi)^{\beta-1}}{{\int_0^1 \Phi^{\alpha-1}(1-\Phi)^{\beta-1}\mathrm{d}\Phi}},
\end{equation}}
where 
\RF{\begin{equation}
	\alpha=\langle\Phi\rangle\left[\frac{\langle\Phi\rangle(1-\langle\Phi\rangle)}{\left\langle\Phi^{\prime} \Phi^{\prime}\right\rangle}-1\right] \quad \mathrm{and} {\quad \beta=\alpha \cdot \frac{1-\langle\Phi\rangle}{\langle\Phi\rangle},} 
\end{equation}}
\RF{with taking $\Phi = \phi/\phi_0$ for concision in equations. The $\beta$-PDF is widely used in studies of turbulent mixing \cite{Madnia1991CEC, Pan2001AIChE, Liu2017AIChE}.} 
\begin{figure*}
	\centering
	\includegraphics[trim=0.45cm 0cm 0.2cm 0.1cm, clip, width=1\textwidth]{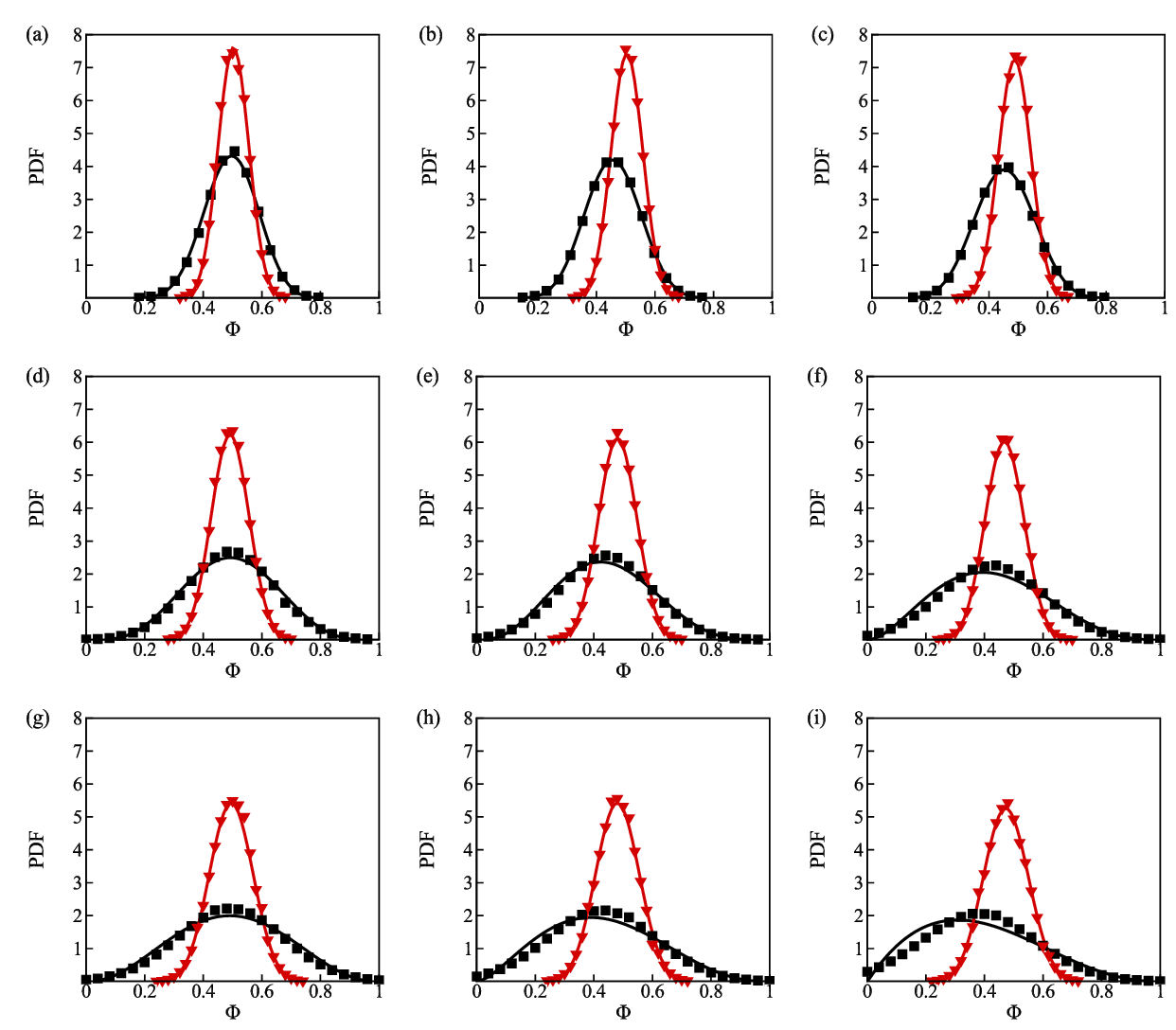}
	\caption{\RF{(Color online) Probability distribution function (PDF) of the scalar from measurements (symbols) and $\beta$-PDF model (lines) at $Re = 1500$ (\textcolor{black}{$\blacksquare$, \sampleline{}}) and $5000$ (\textcolor{red}{$\blacktriangledown$}, \textcolor{red}{\sampleline{}}) at selected points {in the $x-y$ plane ($z/H = 0$)}. From left to right $y/H= 0$, $ -0.5$ and $-0.75$ and from top to bottom $x/H=16$, $12$ and $8$. }}\label{fig:pointPDF}
\end{figure*}

We leveraged our experimental measurements of \RF{$\Phi$} to assess the applicability of the presumed $\beta$-PDF in the T-mixer flow. The symbols in figure~\ref{fig:pointPDF} show the measured PDFs at $Re=1500$ and $5000$ at several different locations in the outlet channel. As shown in figure~\ref{fig:pointPDF}(a), far downstream in the center of the channel, {$(x/H,y/H,z/H)=(16,0,0)$}, both PDFs are centered around \RF{$\Phi = 0.5$}, as expected. Moving toward the wall, as shown in figure~\ref{fig:pointPDF}(b)--(c), the distributions become slightly broader and displaced toward the left, especially for $Re=1500$, where the mixing is much less complete. The presumed $\beta$-PDFs (shown as solid lines) excellently approximate all measured PDFs. Further upstream ($x/H=12$ and $x/H=8$, shown in the second and third rows of figure~\ref{fig:pointPDF}, respectively), the results for $Re=1500$ and $5000$ increasingly diverge. Specifically, the PDFs for $Re=1500$ become increasingly asymmetric far from the channel centre. They are generally much broader and indicate that some fluid parcels are unmixed (see panel i). The presumed $\beta$-PDFs approximate the measured data at $Re=5000$ very well and fairly well for $Re=1500$. 

Generally, more asymmetric PDFs are less well captured. In all cases, the distributions are unimodal and smooth, which suggests that there are no large-scale mixing structures in the present measurements and that turbulent diffusion dominates the mixing in the outlet channel. Overall, the performance of the presumed $\beta$-PDF model appears robust, as demonstrated in \citet{Pan2001AIChE} and \citet{Liu2017AIChE} for other flows. However, it is important to note that the current validation of the model is based on data collected relatively far downstream from the junction. The model's performance closer to the junction, where the mixing dynamics may differ significantly, requires further investigation to assess its applicability.

\subsection{Mechanical-to-scalar timescale ratio}\label{sec:cphi}

Developing a reliable second-order closure for scalar transport requires understanding or calculating the rate at which fluctuations in the scalar field are dissipated by molecular diffusion within fine-scale motions \cite{Beguier1978POF}. This dissipation rate is crucial for formulating closed transport equations, as the ratio of scalar fluctuation levels to their dissipation rate defines the key timescale for characterizing the scalar turbulent field \cite{Corrsion1964, Beguier1978POF}. Most studies on closures assume that the timescale of scalar turbulence ($t_\phi = 2\left<\phi^{\prime}\phi^{\prime}\right>/\epsilon_\phi$) is proportional to the timescale of the turbulent velocity field ($t=k/\epsilon$) \cite{Fox_2003, Pope_2000},
\begin{equation}\label{eq:cphi}
	{C_\phi} = \frac{\epsilon_\phi k}{\epsilon \left<\phi^{\prime}\phi^{\prime}\right>}.
\end{equation}
where the presumed proportionality constant $C_\phi$ is the mechanical-to-scalar time-scale ratio. However, in practice $C_\phi$ is not constant, but depends on the turbulent Reynolds number \cite{Pope_2000, Fox_2003}, {$Re_T={k}/{(\epsilon \nu)^{1/2}}$}.

In the literature, a few models have been proposed to approximate $C_\phi$. For example, \citet{Liu2006AIChE} proposed the empirical equation
\begin{equation}
	C_\phi=\sum_{n=0}^6 a_n\left(\log _{10} R e_T\right)^n,
\end{equation}
for mixing in a confined impinging-jet reactor with $Sc=1000$. However, they did not specify how their reported values of the coefficients  ($a_0=0.4093$, $a_1=0.6015$, $a_2=0.5851$, $a_3=0.09472$, $a_4=-0.3903$, $a_5=0.1461, a_6=-0.01604$) were determined {\citep{Liu2006AIChE}}. For homogeneous isotropic turbulence, \citet{Borgas2004POF} developed a model based on the scalar variance spectrum, 
\begin{equation}\label{eq:Borgas}
	C_\phi = \frac{3}{2C_{Re} - {3}C_B Re_T^{-1} + 2C_B Re_T^{-1} \text{ln}(C_B/C_{OC}^{3/2}) + C_B Re_T^{-1} \text{ln}(Sc)},
\end{equation}
where $C_{Re}$ is a fitted value for the large Reynolds number limit, $C_{OC}$ is the Obukhov--Corrsin constant and $C_{B}$ is the Batchelor constant. These three coefficients can be determined from the scalar spectrum in inertial-convection range, {$E_\phi = C_{OC} \epsilon_{\phi} \epsilon^{-1/3} \kappa^{-5/3}$} and in viscous-convection range, {$E_\phi = C_{B} \epsilon_{\phi} {\left(\nu/\epsilon\right)}^{1/2} \kappa^{-1}$}. Leveraging DNS, \citet{Borgas2004POF} reported that $C_{Re}\approx0.61$, $C_{OC}\approx0.67$ and $C_{B}\approx5$, as listed in table~\ref{tab:Borgas}. 
%The equation~\ref{eq:Borgas} can be re-organized into
%\begin{equation}\label{eq:cphifit}
%	C_\phi = \frac{1}{\alpha Re_T^{-1}+\beta},
%\end{equation}
%to fit the experimental data and extract $C_{Re}$, $C_{OC}$ and $C_{B}$ from the fitting coefficients $\alpha$ and $\beta$. 

We extracted $C_\phi$ via equation~\ref{eq:cphi} using our velocity and scalar field data averaged over $x/H \in[15.8,16.2]$. The analysis spanned Reynolds numbers $Re\in[1500,5000]$ and the spanwise range $y/H\in[0, 0.4]$, and for each $(y,Re)$ pair, we obtained $Re_T$ and {$C_\phi$}. In figure~\ref{fig:cphi}(a) we show that as $Re_T$ increases, $C_\phi$ increases rapidly and then gradually saturates. The model predictions proposed by \citet{Liu2006AIChE} for the confined impinging-jet reactor and by \citet{Borgas2004POF} for homogeneous isotropic turbulence are also shown in figure~\ref{fig:cphi}(a), respectively. Both models predict an increasing trend of $C_\phi$ with $Re_T$ in qualitative agreement with our measurements.

We note that \citet{Borgas2004POF} proposed their model for homogeneous isotropic turbulence, whereas our case corresponds to decaying turbulence, with substantial shear and anisotropies far from the outlet's centerline. 
We fitted our measurement data with equation~\ref{eq:Borgas} and obtained $C_{Re}\approx0.96$, $C_{OC}\approx0.47$ and $C_{B}\approx0.79$, see table~\ref{tab:Borgas}. The fit is in excellent agreement with the measurement and suggests that $C_\phi$ saturates at about $1.89$ for very large $Re_T$. This is slightly smaller than $C_\phi\approx2$ extrapolated from other flows \cite{Beguier1978POF}. 

In the model proposed by \citet{Borgas2004POF}, the parameter $C_{OC}$ and $C_B$ are the proportionality constants of the scalar variance spectra in the inertial-convective and viscous-convective ranges,  i.e. {$E_\phi = C_{OC} \epsilon_{\phi} \epsilon^{-1/3} \kappa^{-5/3}$ and $E_\phi = C_{B} \epsilon_{\phi} {\left(\nu/\epsilon\right)}^{1/2} \kappa^{-1}$,} respectively. This provides an alternative path to test the robustness of the fitted coefficients. For this purpose, we computed $E_\phi$ from our measurements at $Re_T = 45$. Two scalar spectra (in the $x/H$ direction) are shown in figure~\ref{fig:cphi}(b), one at $y/H=0$ for $Re = 4000$ and another one at $y/H = 0.06$ for $Re = 5000$. To mitigate the effects of measurement noise on the spectrum \citep{miller1991stochastic, Lavertu2008JFM}, a Wiener filter was applied \citep{Wiener1964Book}, as specified in \citet{Press2007Book}. Specifically, the noise spectrum (the fluctuations of the scalar concentration of about 0.01$\phi_0$) was subtracted from the raw signal spectrum to enhance the signal-to-noise ratio \cite{Lavertu2008JFM}. \RF{The finest spatial resolution of the measurements is 31 $\mu$m corresponding to a maximum wavenumber of $ \kappa_\mathrm{max} \approx 32.25 \, \text{mm}^{-1} $. Using the Kolmogorov scale ($\eta_k$) estimated from the dissipation data in figure~\ref{fig:tkeudiss}, the normalized spectral cutoffs are $ \kappa_\mathrm{max} \eta_k \approx 11.61 $ for $Re = 4000$ $(y/H = 0)$ and $ \kappa_\mathrm{max} \eta_k \approx 10.32 $ for $Re = 5000$ $(y/H = 0.06)$, where $\eta_k \approx 0.36 \, \text{mm}$ and $\eta_k \approx 0.32 \, \text{mm}$, respectively. This resolution prohibits fully resolving the smallest scales in the viscous-convective range (e.g., the Batchelor scale $\eta_b \approx 9 \, \mu\text{m} $ at $Sc = 1250$ \citep{Crimaldi2001EF} for $Re = 5000$ at $y/H = 0.06$). Nevertheless, the inertial-convective range remains well-resolved. In figure~\ref{fig:cphi}(b), the scalar spectra show a narrow band of $\kappa^{-5/3}$-scaling in the inertial-convective range and an incipient {$\kappa^{-1}$}-scaling in the viscous-convective range. This agrees with the passive scalar mixing recently studied in DNS of homogeneous isotropic turbulence \cite{sreenivasan2019turbulent}. In order to better resolve the dynamics in the viscous-inertial range, e.g., for better observation of the {$\kappa^{-1}$}-scaling, technical improvements are necessary.} We fitted the spectra in the regions shaded in grey in figure~\ref{fig:cphi}(b) and obtained $C_{Re}\approx0.79$, $C_{OC}\approx0.46$ and $C_{B}\approx1.01$, which is in excellent agreement with the fit using equation~\ref{eq:Borgas}. Our data hence support the validity of the model proposed by \citet{Borgas2004POF}, but show that the model's coefficients are flow specific. 

\begin{table}[]
	\centering\caption{Model parameters for the mechanical-to-scalar timescale ratio, $C_\phi$.}\label{tab:Borgas}
	\begin{tabular}{cccc}
		\hline
		Flow & $C_{Re}$ & $C_{OC}$ & $C_{B}$ \\ \hline
		Homogeneous isotropic turbulence \cite{Borgas2004POF} & 0.61 & 0.67 & 5  \\
		T-mixer, equation~\eqref{eq:Borgas} & 0.79 & 0.47 & 0.98 \\ 
		T-mixer, $E_\phi$ spectrum & 0.79 & 0.46 & 1.01 \\ \hline
	\end{tabular}
\end{table}
\begin{figure*}
	\centering
	\includegraphics[trim=0.0cm 0.2cm 0.2cm 0.2cm, clip, width=0.9\textwidth]{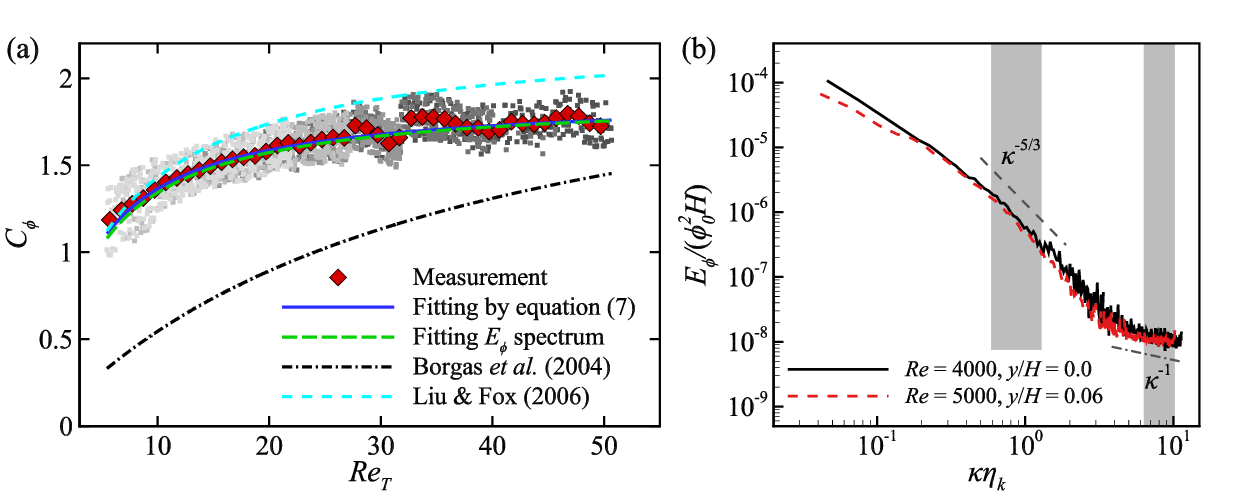}
	\caption{\RF{(Color online) (a) Dependence of $C_\phi$ on the turbulent Reynolds number $Re_T$ extracted from our measurements, with $Re\in[1500,5000]$ and $y/H\in[0,0.4]$. The gray square symbols represent the measured $C_\phi$ with color gradation (light to dark) from low to high $Re$. Red diamond symbols denote averages of the individual points within bins of size of 0.5 in $Re_T$. (b) One-dimensional longitudinal scalar spectra ($x/H\in[15.8,16.2]$) using a Wiener filter for $Re_T = 45$, $(Re, y/H) = (4000, 0)$ (black solid line) and $(Re, y/H) = (5000, 0.06)$ (red dashed line). Wavenumbers are normalized by their respective Kolmogorov scales ($\eta_k$). The gray bands mark the inertial-convective range and the viscous-convective range, respectively, used for coefficient fitting.}}\label{fig:cphi}
\end{figure*}

\section{Conclusions} \label{sec:conclusion}

We designed an upscaled T-mixer to investigate passive-scalar mixing in liquids over a wide range of Reynolds numbers, from laminar to fully turbulent flows. Our PIV measurements at $Re<500$ reproduced the flow regimes in the T-junction reported experimentally and numerically in previous works and validated our setup qualitatively and quantitatively. We investigated the turbulent regime up to $Re=5000$ and showed that the turbulent kinetic energy and turbulent dissipation rate decay consistently with power-laws along the streamwise direction. The scalings appear robust with respect to the spanwise location of the measurement. Our PLIF measurements in the same regime exhibit also a power-like decay of the scalar variance and dissipation, however much more slowly. This is expected from the large Schmidt number of our experiments.  Further, we examined the PDFs of the scalar at different locations in the main channel and found that they are well captured by the presumed $\beta$-PDF model, especially at large Reynolds number, $Re=5000$. At a low Reynolds number, and especially closer to the junction, the distributions become increasingly asymmetric and the $\beta$-PDF model does not work very well, possibly because of the imprints of the large-scale turbulent structures that form close to the junction.  

A crucial dimensionless parameter in the modeling of turbulent mixing is the mechanical-to-scalar ratio, $C_\phi$, because it allows to infer mixing directly from the turbulent kinetic energy and dissipation. Our data support the validity of the model proposed by \citet{Borgas2004POF} for homogeneous isotropic turbulence and show that the model's coefficients are flow specific. We first obtained the coefficients from a fit to the measured $C_\phi$ and also independently from fits to the scalar spectra in the inertial and viscous-convective ranges. Both cases resulted in the same coefficients. Scalar spectra measured at a different Reynolds number and position, but with same turbulent Reynolds number, $Re_T$, were indistinguishable, suggesting the universality of the mixing processes in the outlet channel far from the junction. These spectra showed an incipient $\kappa^{-1}$-scaling measured here in space directly, i.e.\ without using Taylor's frozen field hypothesis.

While these results are encouraging, several important points remain open. First, the inlets in our T-mixer were designed to allow the laminar flow to fully developed. For $Re > 1500$, it appears that the flow started to develop growing disturbances, if not turbulence, near the junction. In future studies it would be desirable to measure the inlet flow conditions simultaneously with the velocity and scalar fields at the outlet in order to better understand discrepancies with direct numerical simulations, as observed in the kinetic energy and dissipation in the central region of the outlet channel. Further refinements of the optical setup would contribute to obtain more refined spectra down to the Batchelor scale and to further probe the power-law decay of the velocity and scalar fields. Indeed, our data show that in the log-log scale the slopes of decay are not quite constant. The corresponding fitted exponents do not yield a constant $C_\phi$, as found at large $Re_T$. Finally, simultaneous measurements of the velocity and scalar field would allow a quantification of the transport equation for the scalar variance. These points will be addressed in future work using the setup introduced here and direct numerical simulations with the ultimate goal of providing full spatio-temporal information of the mixing process and aid the validation and development of models for small-scale mixing.

\begin{acknowledgments}
This work was supported by the Deutsche Forschungsgemeinschaft (DFG, German Science Foundation) with the grant number 394584981 for instrumentation, the grant number 511099203 for the research and {the grant number 34955802 (within the research unit FOR 2688 ``Instabilities, Bifurcations and Migration in Pulsatile Flows")}. H. Li and D. Xu were partially supported by the National Natural Science Foundation of China (Nos. 92152106 and 2372224) and the China Postdoctoral Science Foundation (Nos. 2023M733589 and 2025T180520). {K. Avila acknowledges support from the Ministry of Science and Culture of Lower Saxony through the ``Zukunftskonzept Windenergieforschung"}. The authors thank Katja Kr{\"o}mer, Holger Faust and Peter Prengel for their technical support. 
\end{acknowledgments}

\appendix

\section*{Declarations}
The authors declare that they have no known competing financial interests or personal relationships that could have appeared to influence the work reported in this paper.

%\section{Appendixes}

%\nocite{*}

\bibliography{mixer}
\end{document}